\documentclass{nature}
\usepackage{bm}
\usepackage{amssymb}
\usepackage{epsfig}
\usepackage{color}
\usepackage{amssymb,amsmath}
\usepackage{aas_macros}
\usepackage{graphicx}

\newcommand{\nw}{nW~m$^{-2}$~sr$^{-1}$}

\newcommand{\mll}{$M_{\ell\ell'}$}

\newcommand{\cl}{$C_{\ell}$}
\newcommand{\be}{\begin{equation}}
\newcommand{\ee}{\end{equation}}
\newcommand{\ba}{\begin{eqnarray}}
\newcommand{\ea}{\end{eqnarray}}
\newcommand{\erf}{\mathrm{erf}}

\title{Ultraviolet Luminosity Density of the Universe During the Epoch of Reionization}

\author{Ketron Mitchell-Wynne$^1$, Asantha Cooray$^{*,1}$, Yan Gong$^{2,1}$, Matthew Ashby$^{3}$, Timothy Dolch$^{4}$, 
Henry Ferguson$^{5}$, Steven Finkelstein$^{6}$, Norman Grogin$^{5}$, Dale Kocevski$^{7}$, Anton Koekemoer$^{5}$, 
Joel Primack$^{8}$, Joseph Smidt$^{9}$}

\begin{document}

\maketitle

\begin{affiliations}
\item Dept. of Physics \& Astronomy, University of California, Irvine, CA 92697, USA
\item National Astronomical Observatories, Chinese Academy of Sciences, 20A Datun Road, Chaoyang District, Beijing 100012, China
\item Harvard-Smithsonian Center for Astrophysics, 60 Garden St., Cambridge, MA 02138, USA
\item Department of Astronomy, Cornell University, Ithaca, NY 14853
\item Space Telescope Science Institute, 3700 San Martin Dr., Baltimore, MD 21218
\item Department of Astronomy, The University of Texas at Austin, Austin, TX 78712
\item Department of Physics and Astronomy, University of Kentucky, Lexington, KY 40506, USA
\item Physics Department, University of California Santa Cruz, Santa Cruz, CA 95064, USA
\item Theoretical Division, Los Alamos National Laboratory, Los Alamos, NM 87545, USA
\end{affiliations}


\begin{abstract}
The spatial fluctuations of the extragalactic background light trace the total emission from all stars and galaxies in the Universe. A multi-wavelength study can be used to measure the integrated emission from first galaxies during reionization when the Universe was about 500 million years old. Here we report arcminute-scale spatial fluctuations in one of the deepest sky surveys with the Hubble Space Telescope in five wavebands between 0.6 and 1.6 $\mathbf{\mu}$m. We model-fit the angular power spectra of intensity fluctuation measurements to find the ultraviolet luminosity density of galaxies at $\mathbf{z > 8}$ to be $\mathbf{\log \rho_{\rm UV} = 27.4^{+0.2}_{-1.2}}$ erg s$^{-1}$ Hz$^{-1}$ Mpc$^{-3}$ $\mathbf{(1\sigma)}$. This level of integrated light emission allows for a significant surface density of fainter primeval galaxies that are below the point source detection level in current surveys.
\end{abstract}

\vspace{1cm}
\noindent{\Large {\bf Introduction}}\\
The formation and early evolution of the first galaxies in the universe occurred some time after the dark ages, when the 
coalescence of gravitationally bound masses formed in complex structures, with a spatial distribution that can be traced back to 
primordial overdensities\cite{Loeb2001,Fan2006}. The ultraviolet (UV) photons from these first sources initiated the 
reionization of the surrounding neutral medium, thus ending the dark ages and beginning the era of a transparent cosmos,
which we are increasingly familiar with today.
The luminosity per unit volume of these UV photons at a rest wavelength around 1500\AA\ ($\rho_{\rm UV}$) during this reionization 
period is an important quantity to measure, as it traces the star formation and evolution of these ionizing sources.
The traditional method to measure the UV luminosity density of the universe, $\rho_{\rm UV}$, during the epoch
of reionization, involves searching for candidate galaxies at $z > 6$ through their Lyman-dropout 
signature\cite{Oesch2014,Zheng2012,Coe2013,Bouwens2014b, Finkelstein2014} and then constructing
the luminosity function of those detected galaxies based on the observed number counts. This luminosity function
is then extraploated to a fainter absolute magnitude and integrated in luminosity to calculate $\rho_{\rm UV}$.

There is a second way to quantify  $\rho_{\rm UV}$. This involves a measurement of the extragalactic background light (EBL) 
and, in particular, the angular power spectrum of the EBL intensity fluctuations.
Because these intensity fluctuations are the result of emissions throughout the cosmic time, 
the signal we measure today is the sum of many different emission components, from nearby in our Galaxy to distant sources.
If the integrated intensity from reionization can be reliably separated from that of foreground signals,
we may be able to make an accounting of the total luminosity density of UV photons from reionization.
Just as Lyman-dropout galaxies are detected in deep sky surveys, there is a way to achieve such a separation. 
Due to redshifting of the photons arising from sources present during reionization, 
their emission, as seen today, is expected to peak between 0.9 and 1.1 $\mu$m. This assumes that
the reionization occurred around $z~\sim$7 to 9, consistent with optical depth to electron scattering
as measured by Planck\cite{Planck2015}. Due to absorption of ionizing UV photons, there is no 
contribution shortward of the redshifted Lyman
break around 0.8~$\mu$m\cite{Santos2002, Salvaterra2003}. Spatial fluctuations of the EBL centered around 1~$\mu$m thus 
provide the best mechanism to discriminate the signal generated by galaxies present during reionization\cite{Cooray2004,Fernandez2010} 
from those at lower redshifts, based on the strength of the drop-out signature in the fluctuations measured in different bands.

There are existing measurements of the EBL flcutuations though their origin remain uncertain. This is mostly due to the fact that the
previous measurements  of  EBL fluctuations have until now been limited to wavelengths greater than $1.1 \,\mu$m, with the 
best measurements performed at $3.6 \,\mu$m\cite{Kashlinsky2012, Matsumoto2011, Cooray2012b, Zemcov2014, Seo2015}.
These studies have been interpreted with models involving populations of sources present during reionization at 
redshift $z >8$, direct collapse and other primordial blackholes at $z > 12$ (Ref.~\cite{Cappelluti2013,Yue2013a}),
and with stellar emission from tidally stripped intergalactic stars residing in dark matter halos, or the ``intra-halo light'' 
(IHL)\cite{Cooray2012b} at $z < 3$.  The IHL is diffuse stars in dark matter halos due to galaxy mergers and tidal interactions.
While the relative strengths of these various contributions are still unknown, we expect the signal from
high-redshift galaxies to be separable from low-redshift contributions, including those from faint 
nearby dwarf galaxies\cite{Helgason12}, through a multi-wavelength fluctuation study spanning the 1~$\mu$m range,
including in the optical ($\lambda~\lesssim$~1~$\mu$m) and near-IR ($\lambda~\gtrsim$~1~$\mu$m) wavelengths.

Here we present results from a multi-wavelength fluctuation study using data from the Hubble Space Telescope (HST)
that span across the interesting wavelength range centered at 1~$\mu$m. Through models for multiple
sources of intensity fluctuations, from diffuse Galactic Light to primordial faint galaxies,
we are able to describe the five-band fluctuation measurements in optical and near-IR wavelengths to
obtain a constraint on the UV luminosity density of galaxies present during reionization at a redshift above 8.
We compare our measurement to existing constraints on the quantity and we also discuss implication of our measurement.

\noindent{\Large {\bf Results}}\\
\noindent{\bf Fluctuation Power Spectra.} 
We make use of imaging data from the Cosmic 
Assembly Near-Infrared Deep Extragalactic Legacy Survey\cite{Grogin2011, Koekemoer2011} (CANDELS), a legacy 
program of the Hubble Space Telescope (HST; see Figure 1). Due to extensive data in the Hubble archive, we selected the
southern area of the Great Observatories Origins Deep Survey (GOODS)\cite{Giavalisco2004,Windhorst2011} 
for the measurements (see the Methods section for details on our field selection process). 
This field contains HST observations with two instruments (Wide Field Camera 3 and Advanced Camera for Surveys) that
have the deepest and most continuous coverage. Our total dataset is comprised of observations that
were taken between 2002 and 2012, with exposure times ranging between 180 seconds and 1469 
seconds per frame. We avoided the background gradients evident in the publicly available WFC3 and ACS mosaics by creating our own 
self-calibrated mosaics using custom software\cite{Fixsen2000}, to produce 120 square arcminute mosaics
combined from 234 to 428 (depending on the passband) individual flux-calibrated, flat-fielded (FLT) frames (Figure 2). 
These mosaics are publicly available (see Supplementary Information Note 1).
We mask stars and galaxies using an internally developed masking algorithm, facilitated
by a public multi-wavelength catalog spanning from the ultraviolet to the mid-infrared\cite{Guo2013}. 
The auto- (Figure 3) and cross-power spectra (Figure 4) are computed using standard Fourier Transform 
techniques on the masked images, which retain 53\% of their pixels after masking. A number of corrections are 
performed on the power spectra. Details regarding these corrections can be found in the Methods section. 

Our measurements continue to show the significant excess in the fluctuation
amplitude at 30 arcsecond and larger angular scales, when compared to the
clustering of faint, low-redshift galaxies\cite{Helgason12}.
The large-scale fluctuations correlate between filters (Figure~4). The excess in the amplitude 
of fluctuations relative to faint low-redshift galaxies is consistent with previous 
measurements at 3.6 $\mu$m\cite{Kashlinsky2012,Cooray2012b}.
Our HST-based power spectra probe deeper into the fluctuations and have shapes departing from the 
fluctuations measured with the CIBER sounding rocket experiment\cite{Zemcov2014}.
Due to the shallow depth of the CIBER imaging data, the measured fluctuations there are dominated by the 
shot noise of the residual galaxies at the arcminute angular scales that we probe here with Hubble data (Figure 5).

At angular scales of tens of arcminutes and above, CIBER detected an up-turn in the fluctuations 
with an amplitude well above the level expected from instrumental systematics and residual flat-field 
errors\cite{Zemcov2014}. However, as shown in Figure~5, the combination of CIBER and Hubble fluctuations
is consistent with a power-law clustering signal out to the largest angular scales probed by CIBER.
If the power-law signal is of the form $C_l \propto l^\alpha$, the best-fit slope to combined CIBER and Hubble measurements at 1.6 $\mu$m 
is $\alpha={-3.05\pm 0.07}$. This slope is consistent with Galactic dust, which in emission at 100 $\mu$m 
has a power-law of $-2.89 \pm 0.22$ (Ref.~\cite{Amblard2011}).   At the largest angular scales we could be 
detecting  interstellar light scattered off of Galactic dust or diffuse Galactic light (DGL).
The overall amplitude of fluctuations we measure at 1.6 $\mu$m is consistent with 10\% DGL fluctuations at tens of
arcminute angular scales, given the 100 $\mu$m surface brightness of GOODS-S of $\sim$~0.5 MJy/sr, and existing DGL intensity
measurements (Figure 6). Future measurements in the optical wavelengths over a wider area are necessary to confirm
the Galactic nature of fluctuations at angular scales greater than one degree.

\noindent{\bf Multi-Component Model.} 
For our theoretical interpretation, we invoke a model which involves four main components: (a) intra-halo light (IHL) following
Ref.~\cite{Cooray2012b},  (b) diffuse galactic light (DGL) due to interstellar dust-scattered light in our Galaxy, (c) low-redshift 
residual faint galaxies\cite{Helgason12}; and (d) the high-redshift signal. We assume the flat $\Lambda$CDM model with $\Omega_M=0.27$, 
$\Omega_b=0.046$, $\sigma_8=0.81$, $n_s=0.96$ and $h=0.71$ in our theoretical modeling\cite{Komatsu11}.
We summarize the basic ingredients of our model now while providing references for further details.

For IHL, we follow the model developed in Ref.~\cite{Cooray2012b}.
The mean luminosity of the IHL at rest-frame wavelength $\lambda$ for a  halo with mass $M$ at $z$ is described as
\be \label{eq:L_IHL}
\bar{L}^{\rm IHL}_{\lambda}(M,z) = f_{\rm IHL}(M)L_{\lambda_p}(M)(1+z)^{\alpha}F_{\lambda}(\lambda_0/(1+z)),
\ee
where $\lambda_0=\lambda(1+z)$ is the observed wavelength, $\alpha$ is the power-law index which takes account of the redshift-evolution 
effect, and $f_{\rm IHL}(M)$ is the IHL luminosity fraction of the total halo, which takes the form
\be
f_{\rm IHL}(M) = A_{\rm IHL}\left(\frac{M}{M_0}\right)^{\beta}.
\label{eq:Aihl}
\ee
Here $A_{\rm IHL}$ is the amplitude factor, $M_0=10^{12}\ M_{\odot}$, and $\beta$ is the mass power index. 
In equation~(\ref{eq:L_IHL}), $L_{\lambda_p}(M)=L_0(M)/\lambda_p$ is the total halo luminosity at $2.2$ $\mu$m and at 
$z=0$, where $\lambda_p=2.2$ $\mu$m and $L_0$ is given by\cite{Lin04}
\be
L_0(M) = 5.64\times10^{12}h_{70}^{-2}\left( \frac{M}{2.7\times10^{14}h_{70}^{-1}M_{\odot}}\right)^{0.72} L_{\odot}.
\ee
Here $H_0=70\,h_{70}$ $\rm km\,s^{-1}Mpc^{-1}$ is the present Hubble constant. The $F_{\lambda}$ term is the IHL 
spectral energy distribution (SED), which can transfer $L_{\lambda_p}$ to the other wavelengths and is 
normalized to be 1 at 2.2 $\mu$m (see the discussion in Ref.~\cite{Cooray2012b} for details). We assume the IHL SED to be the same as 
the SED of old elliptical galaxies, which are comprised of old red stars\cite{Krick07}.

The angular power spectrum of IHL fluctuations, $C_{\ell}^{\rm IHL}$, is calculated via a halo model approach\cite{Cooray02} 
and involves both a 1-halo term associated with spatial distribution inside a halo and a 2-halo term involving clustering between halos.
The details are provided in Ref.~\cite{Cooray2012b}. In the clustering calculation,
we assume the IHL density profile follows the Navarro-Frenk-White (NFW) profile\cite{Navarro1997,Cooray2012b}. 
We set the maximum IHL redshift at $z_{\rm max}=6$. 
The $M_{\rm min}$ and $M_{\rm max}$ are fixed to be $10^9$ and $10^{13}$ $M_{\odot}/h$, and the power-law index $\beta$ is fixed to be 
0.1 in this work\cite{Cooray2012b}. The IHL model is then described with two parameters: $A_{\rm IHL}$ in equation~(\ref{eq:Aihl})
and the power-law index $\alpha$ in equation~(\ref{eq:L_IHL}).

A simple test of IHL is to grow the source mask and 
study how the fluctuation power spectrum varies as a function of the mask size.  However, we note that
our model for IHL involves clustering at large angular scales. That is, in our description
IHL is not restricted to regions near galactic disks only. The dependence of the power spectrum
with mask size is studied in Refs.~\cite{Arendt2010,Donnerstein15}. These studies find that 
the fluctuations do not vary strongly with the mask radius, though such studies ignored
the mode-coupling effects associated with the mask as the mask radius is varied. As discussed in Ref.~\cite{Cooray2012b},
in order to test IHL one has to grow the masking area to a factor of 10 larger than the typical mask radius used in the current analyses of fluctuations.
In the case of Spitzer data, where we expect fluctuations to be dominated by IHL, the relatively large 2 arcsecond point spread function 
makes it close to impossible to test IHL directly with a varying mask.
However, additional tests of IHL exist in the literature\cite{Arendt2010}. These include correlations between artificial halos and
masked sources, and correlations between masked sources and foreground galaxies. Such tests have not ruled out the IHL component.
Furthermore, without such a component we are not able to explain the fluctuations measured at wavelengths below 0.8~$\mu$m, as residual 
fiant galaxy clustering\cite{Helgason12} is not adequate to explain the measurements.

The DGL component involves dust-scattered light and it is likely that the same dust is observed by IRAS at far-infrared wavelengths through 
thermal emission.
The DGL component was considered in Ref.\cite{Zemcov2014} and an upper limit on the expected amplitude was included based on
the cross-correlation with 100 $\mu$m IRAS map of the same fields. The CIBER final model results focussing on IHL
to explain the fluctuations did not allow the DGL fluctuations amplitude to vary as a free parameter.
With Hubble data, we find stronger  evidence for DGL or a DGL-like signal once combined with CIBER,
and with an rms amplitude for fluctuations that is at least a factor of 3 larger than the
upper limit used in Ref.\cite{Zemcov2014} based on the cross-correlation with CIBER. 
Moreover, we find that the angular power spectrum is proportional to 
$\sim\ell^{-3}$ over the degree scales measured from CIBER to tens of arcminute scales of CANDELS measurements. 
So we model it with an amplitude factor $A_{\rm DGL}$ as
\be
C_{\ell}^{\rm DGL} = A_{\rm DGL}\ell^{-3}.
\ee
To validate this $\ell^{-3}$ DGL slope dependence, we perform a linear fit in log space at low multipoles of 
the HST 1.6~$\mu$m data simultaneously with the CIBER 1.6~$\mu$m data. We measure a slope of $-3.05 \pm 0.07$, so 
the functional form of our DGL model with $\ell^{-3}$ is appropriate (Figure 5). The power-law behavior of the DGL signal 
is consistent with Galactic dust emission power spectra in far-infrared and sub-mm surveys\cite{Amblard2011}, and dust 
polarization measurements in all-sky experiments like Planck\cite{Planck2014}. We summarize a comparison of our DGL 
intensity measurements to those of existing measurements as a function of wavelength in Figure 5.

The clustering of low-redshift faint galaxies at $z < 5$, where reliable luminosity functions exist in the 
literature, is based on the detailed models in Ref.~\cite{Helgason12}. We follow the calculations presented there to 
establish the expected level of low-redshift clustering, and the uncertainty of that expectation at the 
depth to which we have masked foreground galaxies. Because the low-redshift luminosity functions 
are not steep, unless there is a break or steepening in the luminosity function $-$ which is not supported by 
the halo model $-$ these low-redshift populations do not dominate the clustering we have measured. Given that 
our fluctuation measurements reach the deepest depths provided by both Hubble and Spitzer/IRAC,
it is also unlikely that populations such as extreme red galaxies at $z < 2$ are responsible for the measured fluctuations. 
If there are populations at low redshift responsible for the SED of the fluctuations we measure, they would need to have 
individual SEDs that are consistent with a sharp break redshifted between 0.8 and 1.25 $\mu$m.
While fluctuation measurements in just two bands cannot separate galaxies that have redshifted 4000 \AA\ break, or galaxies that
have redshifted Lyman-$\alpha$ break  between those two bands, with five bands we have adequate knowledge on the SED of fluctuations, and
the shape of the clustering over two decades in angular scales to separate high-$z$ galaxies from low-$z$ faint interlopers. The low-$z$ interlopers are also likely captured
by our IHL model as we cannot distinguish between diffuse stars and faint, dwarf galaxies that happen to be a satellite of a large dark matter halo in our
modeling description.

\noindent{\bf Galaxies During Reionization.} 
The final and critical component in our model is the signal from $z > 6$. We break this signal into two redshift intervals given the placement of
the five ACS and WFC3 bands, based on the Lyman-dropout signal that moves across these bands. In particular, we consider $8 < z < 13$ and $6 < z < 8$
as the two windows. As we discuss later, given the availability of SFR density measurements in the $6 < z < 8$ interval, we mostly allow the signal in that redshift interval
to be constrained by the existing data, and model-fit independently the SFR density in the higher-redshift interval. 
We do not have a strong independent constraint on the $6 < z < 8$ signal since it is only a Lyman-dropout in
our shortest-wavelength band at 0.6 $\mu$m. This allows a better separation of the $8 < z < 13$ from the rest of the signals 
discussed above. To measure $6 < z < 8$ independently, we would need at least one more band below 0.6 $\mu$m.
The signal from $8 < z < 15$ disappears from the three optical bands (0.6 to 0.85 $\mu$m) and is present in the two
IR bands at 1.25 and 1.6 $\mu$m. 

To model the high-redshift signals, we adopt an analytic model\cite{Cooray2012a,Yue2013a} based on the work of Ref.~\cite{Fernandez2012}. It involves
a combination of two separate classifications of stars $-$ moderate-metallicity, second-generation or later stars (PopII), and the first 
generation of stars ever formed in the Universe, hence zero metallicity (PopIII). These are modeled with Salpeter\cite{Salpeter1955} 
and Larson\cite{Larson1999} initial mass functions (IMFs) for PopII and PopIII
stars, respectively. The calculation related to direct stellar emission and the associated nebular lines, including especially
Lyman-$\alpha$ emission, follows the work of Fernandez \& Komatsu\cite{Fernandez2012}. The total integrated intensity from $z_{\rm min} < z < z_{\rm max}$ is
\begin{equation}
\nu I_\nu = \int_{z_{\rm min}}^{z_{\rm max}} dz \frac{c}{H(z)}\frac{\nu(z)\bar{j}_\nu(z)}{(1+z)^2} \, ,
\end{equation}
where $\nu(z)=(1+z)\nu$. The comoving specific emissivity, as a function of the frequency is composed of both PopII and PopIII stars with an assumed z-dependent
fraction as discussed in Ref.~\cite{Cooray2012a} with the form given by
\begin{equation}
f_p(z) = \frac{1}{2} \left[ 1 +\erf\left(\frac{z-10}{\sigma_p}\right)\right] \, ,
\end{equation}
with $\sigma_p=0.5$. The model thus assumes most of the halos have PopIII stars at $z >10$ while PopII stars dominate at redshifts lower than that.

There are a number of theoretical parameters related to this model, especially the
escape fraction of the Lyman-$\alpha$ photons $f_{\rm esc}$, the star-formation efficiency denoting the fraction of the baryons converted to stars in high-redshift dark matter halos, or $f_*$, and the minimum halo mass to host galaxies, or M$_{\rm min}$. 
The overall quality of the data is such that we are not able to independently constrain
all of the parameters related to the high-redshift intensity fluctuation signal. Moreover these parameters are degenerate with each other (i.e.,
changing $f_\star$ can be compensated by a change in  M$_{\rm min}$ for example). Thus given that we do not have the ability to
constrain multiple parameters, we simply model-fit a single parameter $A_{\rm high-z}$ that scales the overall amplitude from the default model, 
interpret that scaling through a variation in $f_*$, and subsequently convert that to a constraint on the SFRD.
We fix our default model to a basic set of parameters, and assume $f_{\rm esc}=0.2$, M$_{\rm min}=5\times10^7$ M$_{\odot}$, and $f_*=0.03$.
The resulting optical depth to reionization of this default model is 0.07, consistent with the optical depth measured by Planck\cite{Planck2015}. 
Among all these parameters, the most significant change (over the angular scales
on which we measure the fluctuations) comes effectively from $f_*$, or the overall normalization of $\bar{j}_\nu(z)$, given that it is directly
proportional to $f_*$. This can in turn be translated to a direct constraint on the SFRD, $\psi(z)$, since with $f_\star$ we are measuring the
integral of the halo mass function such that
\begin{equation}
\psi(z) = f_* \frac{\Omega_b}{\Omega_m} \frac{d}{dt} \int_{M_{\rm min}}^\infty dM  M \frac{dn}{dM}(M,z) \, 
\end{equation}
where $dn/dM$ is the halo mass function\cite{Tinker2008}.

Finally, to calculate the angular power spectrum of fluctuations, we also need to assign galaxies and satellites to dark matter halos.
For that we make use of the halo model\cite{Cooray02}. We make use of the same occupation number distribution as in Ref.~\cite{Cooray2012a}
where the central and satellite galaxies are defined following Ref.\cite{zheng2005}. However, departing from the low-redshift galaxy
models, we take a steep slope for the satellite counts in galaxies with $\alpha_s=1.5$. The low-redshift galaxy clustering and luminosity functions
are consistent with $\alpha_s\sim 1$ (Ref.~\cite{zheng2005}), but such a value does not reproduce the steep faint-end slopes
of the LBG luminosity functions\cite{Oesch2014,Zheng2012,Coe2013,Bouwens2014b, Finkelstein2014}. Such a high slope for the satellites also
boost the non-linear clustering or the 1-halo term of the fluctuations. We do not have the ability to independently constrain the slope of satellites
from our fluctuation measurements. In the future a joint analysis of fluctuations and LBG luminosity functions may provide additional information
on the parameters of the galaxy distribution that is responsible for fluctuations. It may also be that the models can be 
improved with additional external information, such as the optical depth to reionization. 
We also note that other sources at high redshift include direct collapse black holes (DCBHs\cite{Yue2013a}),  but we do not explicitly account for them here as 
the existing DCBH model is finely tuned to match Spitzer fluctuations, and the low signal-to-noise ratio of the Chandra-Spitzer 
cross-correlation results in them residing primarily at $z > 12$. DCBHs at such high redshifts will not contribute to Hubble fluctuations.

Finally, at smaller angular scales, the shot noise dominates the optical and IR background intensity fluctuation. 
Since it is scale-independent, we set it as a free variable parameterized as
\be
C_{\ell}^{\rm shot} = A_{\rm shot} \, .
\ee
This noise term in the fluctuation power spectrum arises because of the Poisson behavior of the galaxies at small angular scales,
a product of the finite number of galaxies. Our measured shot noise comes from a combination of the unmasked, faint low-redshift 
dwarf galaxies, and the high-redshift population.
We do not use the information related to the shot noise in our models but instead treat it as a free independent parameter, 
since we cannot separate the high-redshift shot noise from the shot noise produced by faint, low-redshift dwarf galaxies. 
Here we focus mainly on the clustering at tens of arcseconds and larger angular scales to constrain
SFRD during reionization. In the future, with either a precise model for the
low-redshift galaxies or a model for high-redshift galaxies that determines their expected number
counts as a function of the free parameters such as  M$_{\rm min}$ and  $f_*$, it may be possible to
separate the overall shot noise associated with reionization sources from that of the low-redshift faint galaxies.
If this is the case then it might also be possible to improve the overall constraints on the high-redshift population.
It may also be that, under an improved model, shot noise may end up providing complementary information
to galaxy clustering to break certain degeneracies in model parameters.

Our overall model for the optical and infrared background fluctuations is
\ba
{C_{\ell}} = \left\{\begin{array}{lll} C_{\ell}^{\rm IHL} + C_{\ell}^{\rm DGL} +  C_{\ell}^{\rm low-z}+C_{\ell}^{\rm shot} + C_{\ell}^{6<z<8}+C_{\ell}^{8<z<13} & 
\rm{F125W\,\text{and above} }\\
C_{\ell}^{\rm IHL} + C_{\ell}^{\rm DGL} +  C_{\ell}^{\rm low-z} + C_{\ell}^{\rm shot} + C_{\ell}^{6<z<8}& \rm{F775W\,\text{and}\, F850LP}\\
C_{\ell}^{\rm IHL} + C_{\ell}^{\rm DGL} +  C_{\ell}^{\rm low-z} + C_{\ell}^{\rm shot} & \rm{F606W} \end{array}\right.
\ea
Given that we are not able to constrain the the amplitude of $C_{\ell}^{6<z<8}$ given the degeneracies with the parameters involving the IHL model,
and the fact that we only have a single band below it, we set $C_{\ell}^{6<z<8}$ based on the default prediction of our model, but allow the overall amplitude
$A_{6 < z < 8}$ to vary such that it uniformly samples the 
SFRD between [0.003,0.2] M$_{\odot}$ yr$^{-1}$ Mpc$^{-3}$. The range is fully consistent with the existing measurements on the SFRD 
between $z=6$ and 8 (Ref.~\cite{Bouwens2014b,Zheng2012,Coe2013,Oesch2014}).
Our constraint on $A_{8<z<13}$ is mostly independent of this parameter since we can safely constrain the 
Lyman-dropout signal between 0.8 and 1.25 $\mu$m with our existing data.

We also included the CIBER\cite{Zemcov2014} data at 1.1 and 1.6 $\rm \mu m$ and Spitzer\cite{Cooray2012b} data at 3.6 $\mu$m in our fitting 
process. When compared to the Hubble data at 1.25 and 1.6 $\mu$m, we find the CIBER data are likely dominated by the emission 
from a DGL-like signal at large angular scales, and low-$z$ faint galaxies at $z<5$ at small angular scales (Figure~5). For the fluctuations 
from faint, low-$z$ galaxies, we adopt a model of residual galaxies which is derived from the observations of the luminosity function for different 
near-IR bands\cite{Helgason12}. This model already includes the shot noise term, and we add a scale factor $f_{\mathrm{low}-z}$ to vary the 
low-$z$ angular power spectrum, $C_{\ell}^{\mathrm{low}-z}$, in 1~$\sigma$ uncertainty. For the DGL component, we use the $C_{\ell}^{\rm DGL}$ of 
Hubble data at 1.25 and 1.6 $\mu$m to fit the CIBER data at 1.1 and 1.6 $\mu$m. 

We perform joint fits for Hubble, CIBER and Spitzer data with the Markov Chain Monte Carlo (MCMC) method. The Metropolis-Hastings 
algorithm is used to find the probability of acceptance of a new MCMC chain point\cite{Metropolis53, Hastings70}. We estimate the 
likelihood function as $\mathcal{L}\propto {\rm exp}(-\chi^2/2)$, where $\chi^2$ is given by
\be
\chi^2 = \sum^{N_d}_{i=1} \frac{(C_{\ell}^{\rm obs}-C_{\ell}^{\rm th})^2}{\sigma_{\ell}^2}.
\ee
Here $C_{\ell}^{\rm obs}$ and $C_{\ell}^{\rm th}$ are the observed and theoretical angular power spectra for HST, Spitzer or CIBER data, respectively. 
$\sigma_{\ell}$ is the error for each data point at $\ell$, and $N_d$ is the number of data points. The total $\chi^2$ of HST, Spitzer and CIBER is 
$\chi^2_{\rm tot}=\chi^2_{\rm HST}+\chi^2_{\rm CIBER}+\chi^2_{\rm Spitzer}$. 

We assume a flat prior probability distribution for the free parameters; see Table~1 for prior information.
Both $A_{\rm DGL}$, $C_{\ell}^{\rm shot}$ vary as independent parameters for each band. Both the $A_{\rm DGL}$ and $C_{\ell}^{\rm shot}$ 
parameters are six-fold, with one for each HST band and for Spitzer/IRAC 3.6~$\mu$m. (we combined the two CIBER bands with two of the HST bands).
We have two parameters for IHL and one parameter for the normalization of the reionization galaxies with $A_{8<z<13}$. 
We have two more parameters that we vary, $A_{6<z<8}$ and $f_{\mathrm{low}-z}$. We set a uniform prior on  $A_{6<z<8}$ in the SFRD 
following the existing measurements to be between 0.003 and 0.2 M$_{\odot}$ yr$^{-1}$ Mpc$^{-3}$.
We also set a uniform prior on $f_{\mathrm{low}-z}$ over a reasonable range of models to account for the overall 
uncertainty in the models of Ref.~\cite{Helgason12} to describe the $z < 5$ faint galaxy clustering at the same masking depth as our measurements.
We marginalize over both $A_{6<z<8}$ and  $f_{\mathrm{low}-z}$ as well as all other parameters when quoting results for an individual parameter.
We have a total of 14 free parameters in our MCMC fitting procedure that we extract from the data. 
Among these parameters, 12 of them simply describe the small and large angular scale fluctuations in each of the 
bands we have performed the measurements. These parameters are summarized in Table 1. We generate twenty MCMC chains, where each chain contains about 
100,000 points after convergence. After thinning the chains, we merge all chains and collect about 10,000 points for illustrating 
the probability distributions of the parameters. Contour maps for each of the fitted model parameters are shown in Figure~8.
Our best-fit model with 14 free parameters have a minimum $\chi^2$ value of 278 for a total degrees of freedom of $N_{\rm dof}=104$.

\noindent{\Large {\bf Discussion}}\\
Our results are summarized in Figure~3, where we show the best-fit model curves. 
While the dominant contribution to the excess fluctuations comes from DGL at $\ell < 10^4$, at intermediate scales we find the IHL and 
reionization contributions to be roughly comparable. In Figure~6 we show the rms fluctuation 
amplitude at $\sim$~5 arcminute angular scales over the interval $10000  < \ell < 30000$.
We find a spectral energy distribution that is consistent with Rayleigh-Jeans (RJ) from 
4.5 to 2.4~$\mu$m, but diverges between 2.4 and 1.6~$\mu$m, and even more rapidly between 1.25 
and 0.85~$\mu$m. The fluctuations can be explained with a combination of
IHL and high-redshift galaxies. The residual low-z galaxy signal is small but non-negligible. We find that it
is mostly degenerate with IHL, especially if we allow its amplitude to vary more freely than the range allowed
by the existing models based on $z < 5$ galaxy luminosity functions\cite{Helgason12}. Thus modeling uncertainities related to
the low-$z$ galaxy confusion do not contaminate our statements about reionization. Assuming the existing low-$z$ 
galaxy model\cite{Helgason12}, the best-fit model is such that the IHL intensity peaks at lower redshifts with decreasing 
wavelength (Figure~7). At 3.6 $\mu$m, the IHL signal is associated with 
galaxies at $z \sim 1$, while at 0.6 $\mu$m over 80\% of the signal is associated with galaxies at
$z < 0.5$. The total intensities are $0.13^{+0.08}_{-0.05}$, $0.23^{+0.17}_{-0.11}$, $0.27^{+0.21}_{-0.13}$,
$0.45^{+0.43}_{-0.24}$, and $0.54^{+0.58}_{-0.31}$ \nw~ at 0.60, 0.77, 0.85, 1.25 and 1.6 $\mu$m, 
respectively.  We  find that the implied IHL intensities at 1.25 and 1.6 $\mu$m are a factor of 10 lower than the
implied IHL intensities for a model of CIBER fluctuations with IHL alone. The difference is due to the
CIBER model that only included IHL and ignored the presence of DGL.

The drop in the fluctuation amplitude from $1.25$/1.6~$\mu$m to 0.85~$\mu$m allows for a signal from reionization, 
but the presence of fluctuations at shorter wavelengths, such as 0.6 $\mu$m, rules 
out a scenario in which reionization sources are the sole explanation for the fluctuations at wavelengths at 1~$\mu$m and above.
The 3.6 $\mu$m and X-ray cross-correlation\cite{Cappelluti2013} was  explained with primordial direct collapse blackholes at $z>12$ (Ref.~\cite{Yue2013a}).
In our multi-component model we are able to account for the presence of fluctuations at short wavelengths with IHL, DGL and faint
low-redshift galaxies, while a  combination of those components and high-redshift galaxies are preferred
to account for fluctuations at 1.25 and 1.6 $\mu$m. The high-$z$ signal is modeled following
the calculations in Ref.~\cite{Cooray2012a}. The signal has an overall amplitude scaling that is related to the star-formation
rate during reionization. The bright end of the counts are normalized to existing luminosity function measurements, and 
the faint-end of the luminosity functions to have a steeper slope than measured with counts extending 
down to arbitrarily low luminosities. In order to test whether a component at high redshift is required to explain the measurements,
we also re-ran the MCMC model fits but with $A_{\rm high-z}$ fixed at $0$. 
In this case our best-fit model with 13 free parameters has a minimum $\chi^2$ value of 283 for a total degrees of freedom of $N_{\rm dof}=105$.
The difference in the best-fit $\chi^2$ values with and without a model for high-redshift galaxies suggests a $p$-value of 0.025. 
This is consistent with the 2$\sigma$ to 3$\sigma$ detection of $8 < z< 13$ signal in the fluctuations (Figure~6).

With multi-wavelength measurements extending down to the optical, we are now able to constrain the amplitude of that signal
with a model that also accounts for low-redshift sources in a consistent manner. This improves over
previous qualitative arguments that have been made, 
or models involving high-redshift sources alone that have been presented, for the presence of a signal from 
reionization in the IR background fluctuations\cite{Kashlinsky2012,Matsumoto2011,Yue2013a}.
In our models, the total intensity arising from all galaxies at $z>6$ is $\log \nu I_\nu = -0.32 \pm 0.12$ in units of
\nw~at 1.6 $\mu$m. At 1.6 $\mu$m the intensity from high-redshift sources is dominated by $z > 8$ galaxies, while at 0.85 $\mu$m we find
an intensity $\log \nu I_\nu = -0.75 \pm 0.05$ in units of \nw~for $6<z<8$ galaxies.  
The total intensity from $z > 8$ galaxies in the 1.6 $\mu$m band is comparable to the IHL intensity at the same wavelength (Figure~7). 
However, at 3.6 $\mu$m, the IHL signal is a factor of about 5 times brighter than the $z > 8$ galaxies.  At 1.6 $\mu$m the total of the IHL, high-$z$, and
integrated  galaxy light\cite{Franceschini2008} of $10.0^{+2.7}_{-1.8}$ \nw is comparable to the
EBL intensity inferred by gamma-ray absorption data\cite{Hess2013} of $15 \pm 2 {\rm (stat)} \pm 3 {\rm (sys)}$.

Using the best-fit model and uncertainties as determined by MCMC model fits, we also convert the $A_{8<z<13}$ constraint to
a measure of the luminosity density of the universe at $z > 8$ (Figure~9).
The resulting constraint is $\mathbf{\log \rho_{\rm UV} = 27.4^{+0.2}_{-1.2}}$ in units of erg s$^{-1}$ Hz$^{-1}$ Mpc$^{-3}$ at $(1\,\sigma)$.
As shown in Figure~9, the 68\% confidence level constraint on $\rho_{\rm UV}$ is higher than the existing results from
Lyman drop-out galaxy surveys during reionization at $z > 8$ (Refs.\cite{Atek2015, Robertson2013}), and especially at $z \sim 10$ (Ref.~\cite{Zheng2012}).  
At the 95\% confidence level, our measurement is fully consistent with the existing results at $z \sim 10$ (Ref.~\cite{Bouwens2015}).
Our constraint allows for the possibility that a substantial fraction of the UV photons from the reionization era is coming from
fainter sources at depths well below the detection threshold of existing Lyman dropout surveys, as is indeed 
anticipated from the steep measured slopes of the UV luminosity functions from detected galaxies. Despite their lack of detections in
the deepest surveys with HST, the majority of the faint sources responsible for both fluctuations and
reionization should be detectable in deep surveys with JWST centered at 1~$\mu$m.


\noindent{\Large {\bf Methods}}\\
\noindent{\bf Field selection.} In order to obtain angular power spectra over large angular scales, individual exposures 
must be combined into one or more mosaiced images. 
We generate our own mosaics using the self-calibration technique\cite{Fixsen2000} (SelfCal), instead of using the publicly available 
mosaiced images produced by astrodrizzle (available at http://candels.ucolick.org/data\_access/Latest\_Release.html). 
Foreground emissions, predominately that of Zodiacal light, are particularly pernicious at infrared wavelengths\cite{Kelsall1998}, 
so care must be taken when producing mosaics which combine observations taken at different times, especially at the WFC3/IR channels. Offsets
between frames will lead to a fictitious anisotropy signal if care is not taken to properly model and remove those offsets, which is
what SelfCal was designed to do.

Although the CANDELS observations cover multiple fields, only the two deep fields$-$the Great Observatories Origins 
Deep Survey-South and -North (GOODS-S and GOODS-N)\cite{Giavalisco2004}$-$have sufficient overlap between frames
to perform a self-calibration. The GOODS-N dataset has a larger number of frames with clear overall offsets resulting 
from scattered light than does GOODS-S. Therefore we have restricted our analysis to GOODS-S, which has an area of approximately
120 square arcminutes. The wider CANDELS fields are composed of much poorer tile patterns, as can be seen in Fig. 18 of Ref.~\cite{Grogin2011}, 
which are a significant draw-back for fluctuation studies, since one cannot calibrate the full mosaic to a consistent background 
level without introducing artificial gradients to the background intensity. Such dithering patterns were pursued by CANDELS 
to maximize the total area covered with WFC3. The increase in area is of benefit to studies that aim to detect rare galaxies, 
such as Lyman-break galaxy (LBGs) at $z > 5$.

\noindent{ \bf Initial data reduction and map-making.} In addition to the data collected by CANDELS, the GOODS-S field has a wealth of 
HST archive data, publicly available on the Barbara 
A. Mikulski Archive for Space Telescopes (MAST; located at https://archive.stsci.edu/hst/search.php). 
We assembled our own collection of 
calibrated, flat-fielded (FLT) frames from the MAST archive, comprised of some or all of the data from ten different HST 
proposals\cite{Beckwith2006, Giavalisco2004, Grogin2011, Koekemoer2011}. 
These data are also supplemented by the Early Release Science observations\cite{Windhorst2011} (ERS). The tile patterns
of these observations can be found in Figure~1.

In addition to selecting frames with a favorable tile pattern appropriate for self-calibration, we also had to take
two additional potential issues into account. After the replacement of the ACS CCD Electronics Box during the fourth 
Hubble servicing mission (SM4), ACS imaging data are plagued with horizontal striping dominated by 1/f noise. Furthermore, 
ACS frames have a tendency to introduce a Moir\'{e} pattern (correlated noise) when the pixel scale is modified in a low 
signal-to-noise area. Both of these characteristics can potentially contaminate an angular power spectrum to such an extent 
that the systematics dominate the measurement. With simulations, we found that with an increased number of ACS frames 
taken with varying position angles effectively removes the Moir\'{e} pattern upon repixelization, and the bias-striping issue
is ameliorated by simply omitting a large percentage of post-SM4 frames. Any given collection of ACS frames we used contained
$< 27\%$ post-SM4 frames. 

Our MAST archive data were initially reduced with PyRAF version 2.1.1. MAST queries are reprocessed ``on-the-fly'', which entails using 
the most recent calibration files. Thus the FLT frames we retrieved from the archive had standard calibrations of bias and dark frame 
subtraction, along with flat-field correction, already performed. We identified cosmic rays in the the FLT frames 
with the \textsc{crclean} PyRAF module; sub-arcsecond astrometric alignment against the publicly available CANDELS 
mosaics\footnote{http://candels.ucolick.org/data\_access/GOODS-S.html} was achieved with \textsc{tweakreg}. All ACS data were 
charge transfer efficiency (CTE) corrected, and post-SM4 ACS frames were de-striped prior to CTE correction. 

We generate mosaics from the reduced FLT frames using the same SelfCal model as in Ref.\cite{Arendt2002}, an example 
where HST data has already been self-calibrated; details of the model can be found there. We de-weight bad pixels and cosmic rays, 
and iterate three times in order to find a SelfCal solution. Our input FLT frames are geometrically distorted with a pixel size of 
$0.''0498\,\times\,0.''0502$ for ACS, and $0.''1354\,\times\,0.''1210$ for WFC3. We remove the distortion in the map making procedure 
and produce mosaics with a slightly larger pixel size of $0.''140$ (geometrically square). 

Note that the algorithm we use is same as the one used to generate self-calibrated maps of IRAC in the Spitzer 
fluctuation studies\cite{Kashlinsky2005,Kashlinsky2012,Cooray2012b}. The same method was implemented by the Herschel SPIRE Instrument Science 
team to generate wide area mosaics of the Herschel-SPIRE data, resulting in far-infrared fluctuations\cite{Amblard2011}. 
The algorithm originates from the time of FIRAS\cite{Fixsen2000} and has wide applications. In the future we expect it will be used to combine
frames and produce stable wide-area mosaics from JWST, Euclid, and WFIRST, among others.

We generate two maps per band so we can use the differences and sums to study systematics and noise biases, 
as was done in previous studies\cite{Cooray2012b}. The data are sorted by observation date and every other 
FLT frame was used for each half map. HST data generally have two or more exposures per pointing, 
so this results in two maps per band of the same or similar dither pattern and exposure time per pointing. 
One map from each band can be found in Figure~2.  Multiple maps of the same band enable us to 
do cross-correlations, which ensures a removal of uncorrelated noise in the auto-spectra. 
This jack-knife process is similar to all other analyses related to large-scale structure and CMB 
angular power spectra from maps.

\noindent{\bf Generation of resolved source mask.} We utilized existing multi-wavelength catalogs of 
detected sources from the ultraviolet to mid-infrared 
(CITO/MOSAIC, VLT/VIMOS, VLT/ISAAC, VLT/HAWK-I, and $Spitzer$/IRAC)\cite{Guo2013}. In addition, all the sources from the CANDELS, HUDF 
and ERS surveys (F435W, F606W, F775W, F814W, F850LP, F098M, F105W, F125W, and F160W) are also present in the catalog. The 50\% completeness 
limit for F160W in the catalog is $m_{\text{AB}}\, = 25.9, 26.6$ and $28.1$ for the CANDELS wide, deep and HUDF regions respectively; 
the 5~$\sigma$ limiting magnitudes are  27.4, 28.2 and 29.7.  For each source detected by any of the aforementioned instruments, 
we apply an elliptical mask with parameters corresponding to the \textsc{SExtractor} Kron elliptical aperture. This 
catalog simplifies our masking procedure and ensures we are masking 
sources detected at other wavelengths, which may otherwise be undetected in our five bands. 

In addition to the source mask generated from sources detected by other instruments, we also generate our own internal masks. 
We run \textsc{SExtractor} on a coadded map in each of our five filters and apply the same elliptical masking procedure to 
incorporate the shapes of sources. Next, we take the union of all five internal source masks, plus the source mask we made 
from the pre-existing catalog. After applying this union mask to each band, we clip 5~$\sigma$ outliers and visually 
inspect each masked map. Any residual sources are masked by hand. We verified that all sources detected above 5$\sigma$ 
in any of the bands, including deep IRAC data at 3.6 $\mu$m, are masked. This process yields 53\% of the pixels unmasked 
for the FFT computation. Note that tests can be performed which expand and shrink the source mask to further test
the IHL model (Ref~\cite{Arendt2010}).

\noindent {\bf Absolute flux calibration.} SelfCal achieves relative calibration between frames, so in general, the 
absolute flux calibration, or gain, of SelfCal output maps needs to be determined from a standard flux reference. The multi-wavelength 
catalog used in the masking procedure is in principle a good enough reference, however the photometry in the public CANDELS 
source catalogs was obtained with a private version of \textsc{SExtractor}, and has aperture corrections applied to the 
flux densities which were extracted from PSF-matched maps. We instead generate internal catalogs from public 
CANDELS MultiDrizzled mosaics using the same procedure as we used on our mosaics.

In each band, we re-pixelize the MultiDrizzle maps to our SelfCal pixel scale, and perform source extraction with 
\textsc{SExtractor} on both our mosaics and the 
MultiDrizzle mosaics. We use the same parameter files for all the source extractions. We then astrometrically match the resultant catalogs in each 
band and keep those sources that are common within a radius of 0$''$.1 (our FLT frames were aligned in \textsc{tweakreg} with MultiDrizzle mosaics as 
the astrometric reference, so our astrometry is similar to the MultiDrizzle maps at sub-arcsecond scales). Counts in e$-$/s are converted to $\mu$Jy using 
the current HST magnitude zero points\cite{Koekemoer2011}. The calibration introduces a $4-5$\% error which is propagated into our final error bars.

\noindent{ \bf Power spectrum evaluation.} All the statistical information contained in any one of our maps is summarized by its 
angular power spectrum, $C_{\ell}$, which is just the 
variance of the $a_{\ell m}$'s. We used standard Fast Fourier Transform (FFT) techniques to estimate the $C_{\ell}$'s of our masked
maps\cite{Cooray2012b, Kashlinsky2012, Zemcov2014, Amblard2011}. The mosaics can be seen in Figure~2, both in real space and Fourier space. 
For each band we have two half maps, A and B. To measure the inherent noise 
in the data, we compute the noise power spectrum as the auto power spectrum of (A$-$B)/2.
To measure the raw auto-spectrum, we compute the cross spectrum of the two half maps, A$\times$B, which eliminates any uncorrelated noise in our power 
spectrum estimate. In general, the standard deviation at each multipole is 
\begin{equation}
  \label{eq:cv}
  \delta C_\ell = \sqrt{ \frac{2}{f_{\mathrm{sky}}(2\ell+1)\Delta\ell}}\, (C_\ell^{\mathrm{auto}}+ C_\ell^{\mathrm{noise}}),
\end{equation}
where $\Delta\ell$ is the bin width for the given \cl, and $f_{sky}$ is the fractional sky coverage from all the pixels used in the FFT (excluding zeros). 
In our case, we have some error associated with the absolute calibration of the maps, so we take the total error 
budget of our raw power spectra as the quadratic sum of the calibration errors with the variance in equation~(\ref{eq:cv}). The final measured auto-spectra
and associated errors can be found in Supplementary Table 1.

To account for the effects that the source mask, tile pattern and finite beam size introduce into the power spectrum, we 
employ the correction techniques of the MASTER algorithm\cite{Hivon2002}, and closely follow the implementation procedures explained in 
Section 4, 5 and 6 of the Supplementary Information (SI) of Ref.\cite{Cooray2012b} (Including SI Figure 1 of Ref.\cite{Cooray2012b}). 
Among the procedures listed there, we have only slightly 
modified the way we generate our transfer function, T($\ell$). In addition to adding instrumental noise (step 2
in Section 6 of the SI of 
Ref.\cite{Cooray2012b}), we also add an offset to each tile equal to the median of the given FLT frame. This additional step should 
in principle be a good indicator of how well SelfCal is performing in offset removal, unique to each observation. Transfer functions
for each of our five bands can be found in Supplementary Figure~2; measurements of the beam transfer function can also be found in Supplementary Figure~2.

\noindent {\bf Acknowledgements }This work is based on observations taken by the CANDELS Multi-Cycle
Treasury Program with the NASA/ESA HST, which is operated by the Association of Universities for Research in Astronomy, Inc., under
NASA contract NAS5-26555. A.C. acknowledges support from NSF CAREER AST-06455427, AST-1310310,
and STScI Archival Research program. We thank R. Arendt for useful discussions pertaining to the CANDELS map-making process.

\noindent {\bf Author contributions } K.M.W collected the data from the archive, developed the reduction and analysis pipeline, 
and performed the power spectrum analysis. A.C. developed the model, supervised the research of K.M.W. and Y.G., 
and wrote much of the text in the Article. Y.G. interpreted the power spectrum measurements with models.
M.A., A.C., T.D., H.F., N.G., D.K., A.K. and J.P. are members of the CANDELS project (led by H.F. as a Co-PI) and 
obtained the necessary key data used in the study. J.S. provided suggestions for the power spectrum analysis.
All coauthors provided feedback and comments on the paper.

\noindent {\bf Competing financial interests } The authors declare no competing financial interests. Correspondence and 
requests for materials should be addressed to A.C. (acooray@uci.edu).


\clearpage

\begin{figure}
  \includegraphics{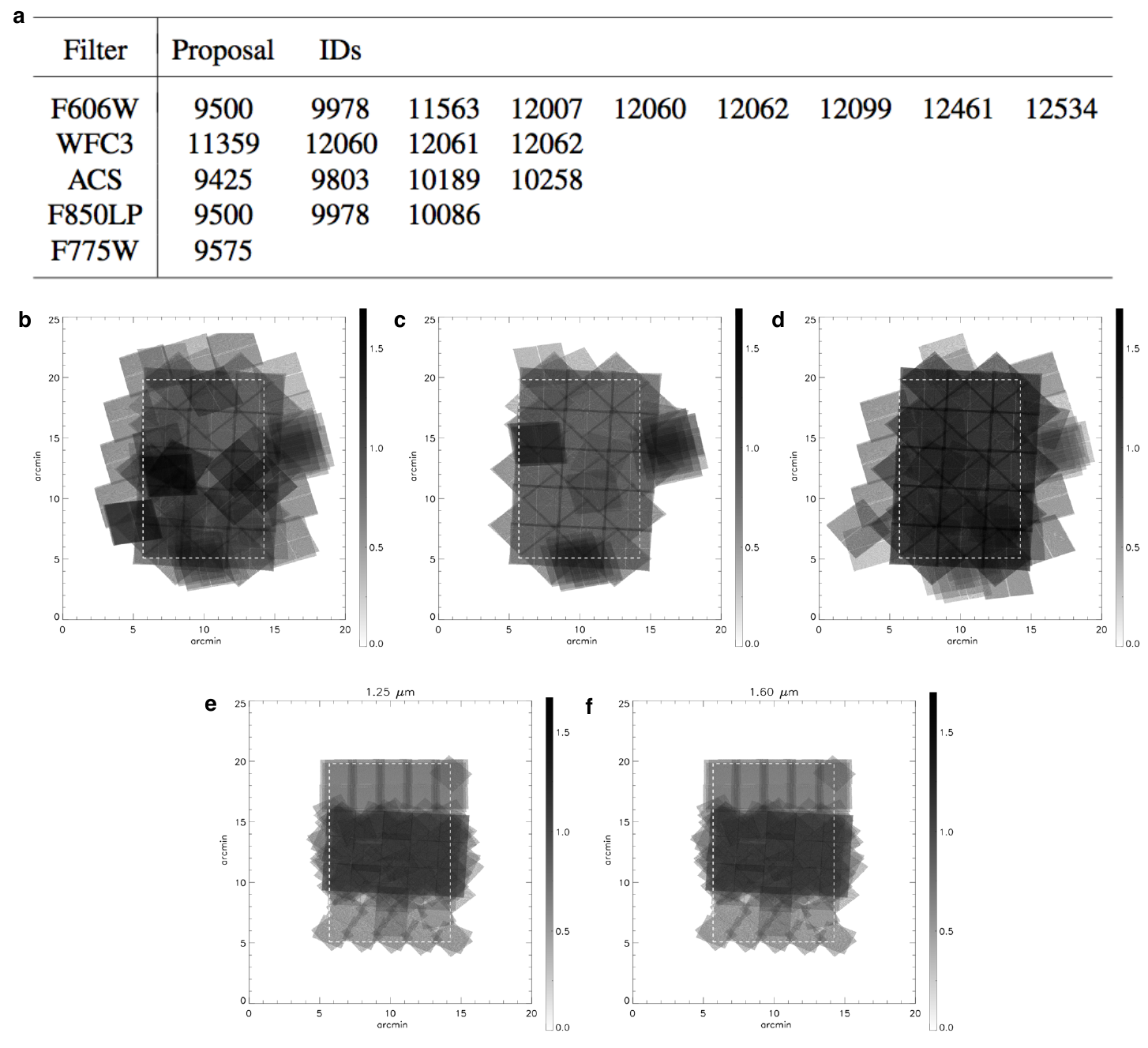}

            {\bf Figure 1 $|$ Summary of tile patterns and their data archive identifications.} {\bf a}, 
      proposal ID's for each filter in the GOODS-S field. 
      The ACS and WFC3 rows show the proposals which are common between all the bands in each instrument. For each 
      proposal we did not necessarily use all the frames, specifically those from deep surveys. Also show are the tiling 
      patterns for all the bands: 0.606~$\mu$m (F606W; {\bf b}), 0.775~$\mu$m (F775W; {\bf c}), 0.850~$\mu$m (F850LP; 
      {\bf d}), 1.25 $\mu$m (F125W; {\bf e}) and 1.60~$\mu$m (F160W; {\bf f}). The units of the tile 
      pattern figures are Log$_{10}$(N+1), where N is the number of frames overlapping. The dashed white line 
      indicates the cropped region where the fluctuation analysis was performed. 

\end{figure}

\clearpage

\begin{figure}
  \includegraphics{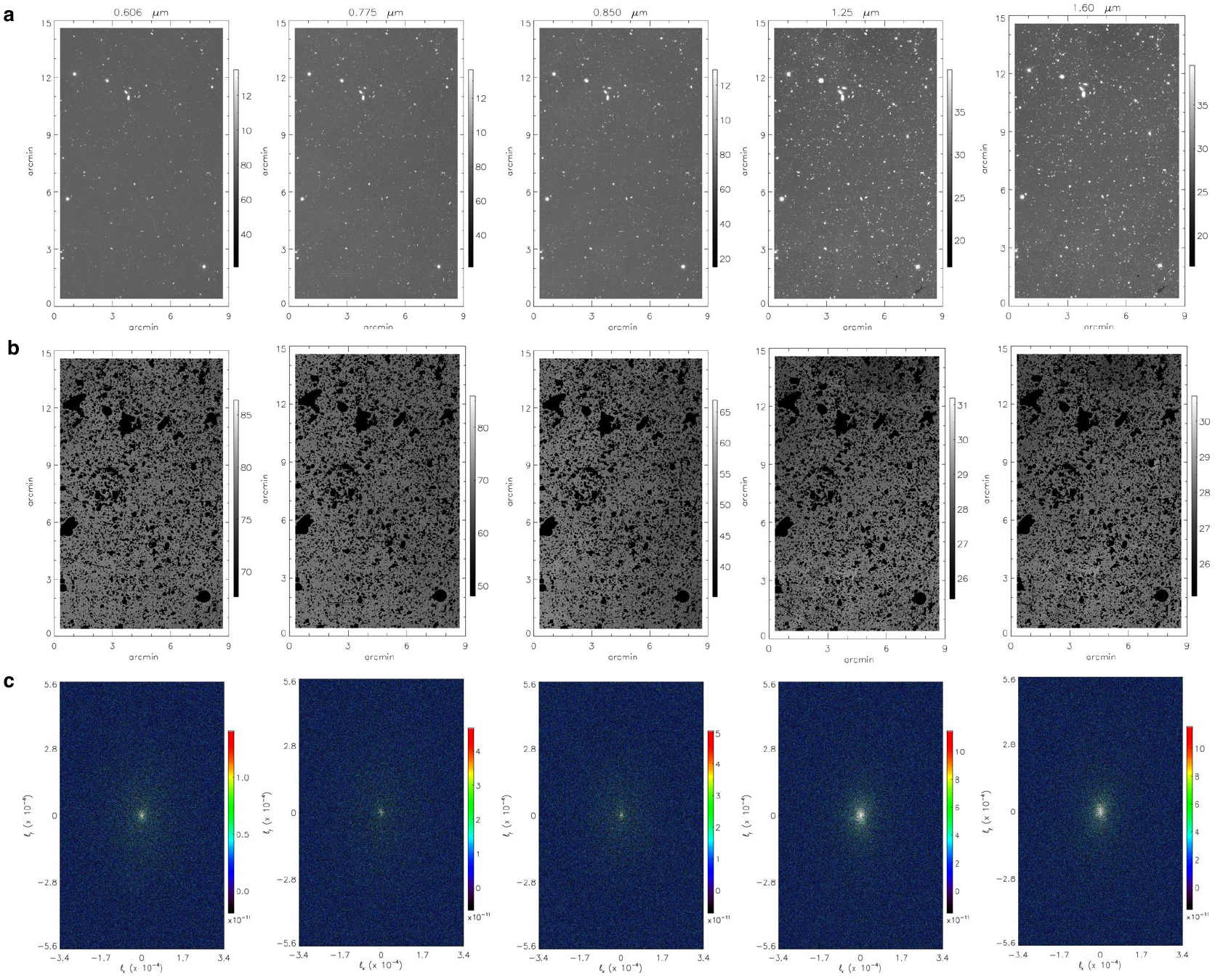}

  {\bf Figure 2 $|$ Self-Calibrated Mosaics. a,}
  GOODS-S SelfCal mosaics for each band. We astrometrically align each map and
  crop the outer regions so as to include only sections that have been observed in all five bands. 
  The units of the maps are ~\nw. {\bf b,} the same as {\bf a} except with the source mask applied. 
  {\bf c} shows the fast fourier transform (FFT) of each of the maps in {\bf b}, which is what is used 
  to measure the angular power spectrum. This is plotted in Fourier space as a 
  function of modes $\ell_{\rm x}$ and $\ell_{\rm y}$. The FFT of each map is structureless and contains 
  only Gaussian noise, which 
  is indicative of high quality mosaics. By definition, the units of the FFT are the same as the units of the map. 
  Each column is filter specific, plotted as 0.606~$\mu$m, 0.775~$\mu$m, 0.850~$\mu$m, 1.25~$\mu$m and 1.60~$\mu$m.  
\end{figure}

\clearpage

\begin{figure}
  \begin{center}
    \includegraphics[scale=1.2]{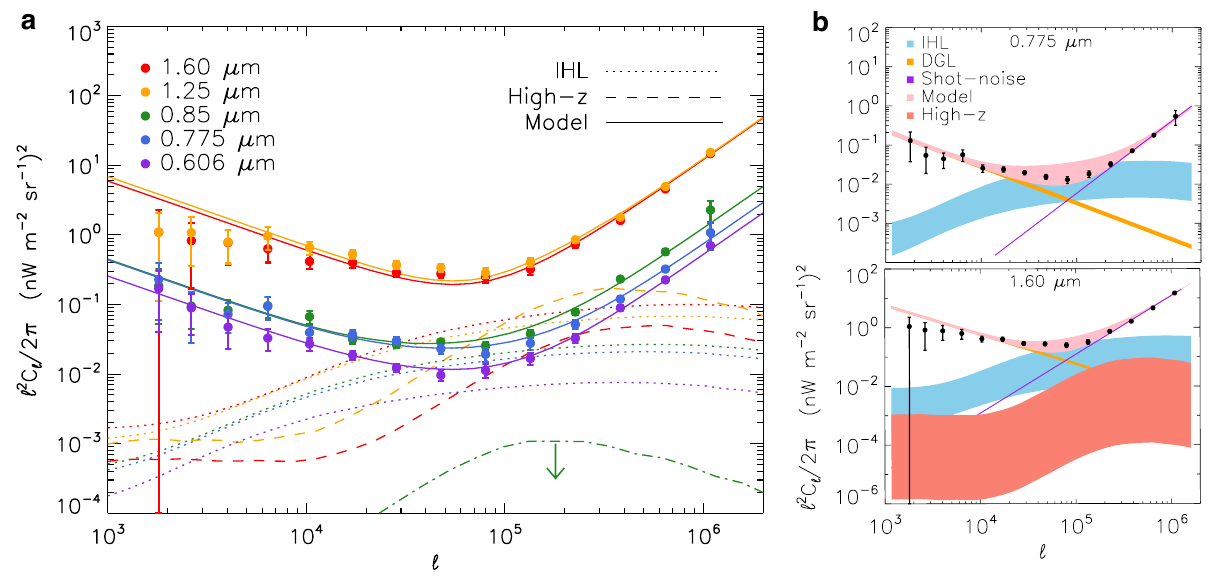}
  \end{center}
  {\bf Figure 3 $|$ Angular power spectra of optical and near-infrared background intensity fluctuations}.
  {\bf a}: Multi-wavelength auto power spectra of optical to near-IR intensity fluctuations in the GOODS-South field 
  using Hubble Space Telescope data (see Online Methods for data selection details). The error bars are calculated by 
  adding in quadrature the errors from the beam transfer function, map-making transfer function and calibration errors, 
  to the standard deviation at each multipole, $\delta C_\ell$, described in Equation~11. Thus the 1~$\sigma$ 
  uncertainties account for all sources of noise and error, including map-making, calibration, detector noise, 
  and cosmic variance associated with finite size of the survey. We show the best-fit model which makes
  use of four components: (a) $z > 8$ high-redshift galaxies, (b) intra-halo light (IHL)\cite{Cooray2012b},
  (c) faint low-redshift galaxies\cite{Helgason12}, and (d) diffuse Galactic light. At 1.25 and 1.6 $\mu$m,
  the best-fit high-redshift galaxy signal is shown as dashed lines. The signal is zero in the optical bands. We show the
  upper limit (denoted by a downard facing arrow) of fluctuations generated by $6 <z < 8$ galaxies as a dot-dashed line. 
  Fluctuation power spectra and the best-fit 
  models with 1$\sigma$ error bounds for the model components are shown at 0.775 $\mu$m in {\bf b} 
  and 1.60~$\mu$m in {\bf c}. The dominant model contributors to the total power spectrum are DGL 
  at low multipoles, or angular scales greater than a few arcminutes, IHL at intermediate multipoles 
  corresponding to angular scales of about an arcminute, and shot noise associated with 
  faint low-redshift dwarf galaxies dominating the high multipoles or sub-arcminute angular scales. 
  The clustering signal of low-$z$ galaxies is more than an order of magnitude below the 
  lower limit plotted here, thus we did not include a low-$z$ component in our modeling.
\end{figure}

\clearpage

\begin{figure}
  \includegraphics[scale=.5]{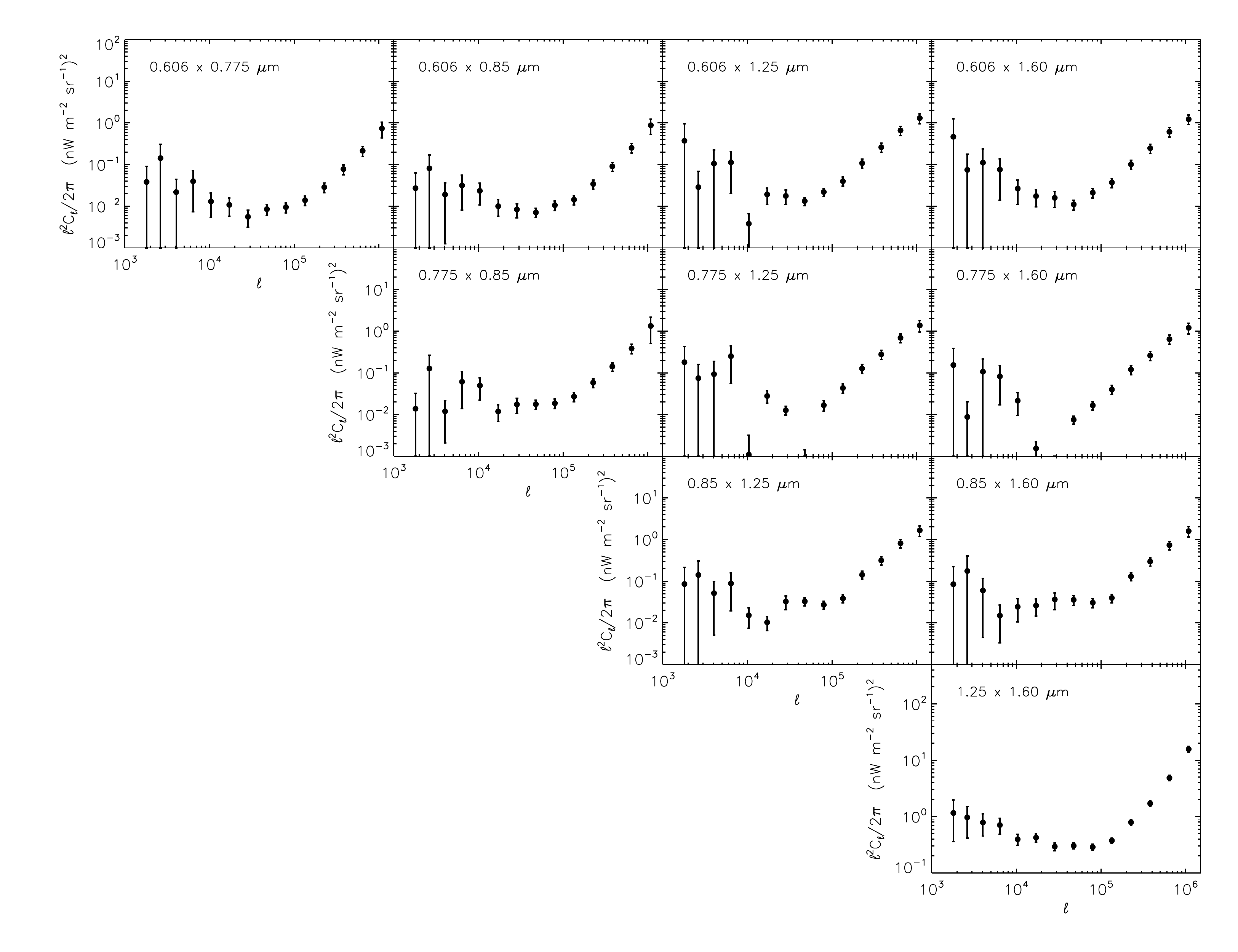}
  {\bf Figure 4 $|$ Angular cross- power spectra of optical and near-infrared background.} Ten cross-correlations between the HST bands. 
  Excess signal is detected in the cross-correlations. The error bars are 1~$\sigma$ uncertainties which are calculated in a similar way as 
  in Figure 3, which accounts for all sources of noise and error, including map-making, calibration, detector noise, and cosmic variance. 
  However, the noise power spectra for the cross-correlations are calculated slightly differently. For each filter we have two maps, so for 
  each cross-correlation between bands we have four maps (label them A and B for the first filter, and C and D for the second). This enables 
  us to generate a noise power spectrum by computing (A-B) x (C-D), as opposed to taking the auto-spectrum of (A-B) for the auto-correlations.
  The first row corresponds to all correlations with the 0.606~$\mu$m band, the second for all correlations with the 0.775~$\mu$m band not found
  in the first row, the third row corresponds to all correlations with the 0.850~$\mu$m band not found in any of the preceding rows, and the last
  row corresponds to correlations at 1.25~$\mu$m. The columns similarly increase in wavelength as you move across the page.
\end{figure}

\clearpage

\begin{figure}
  \includegraphics[scale=.5]{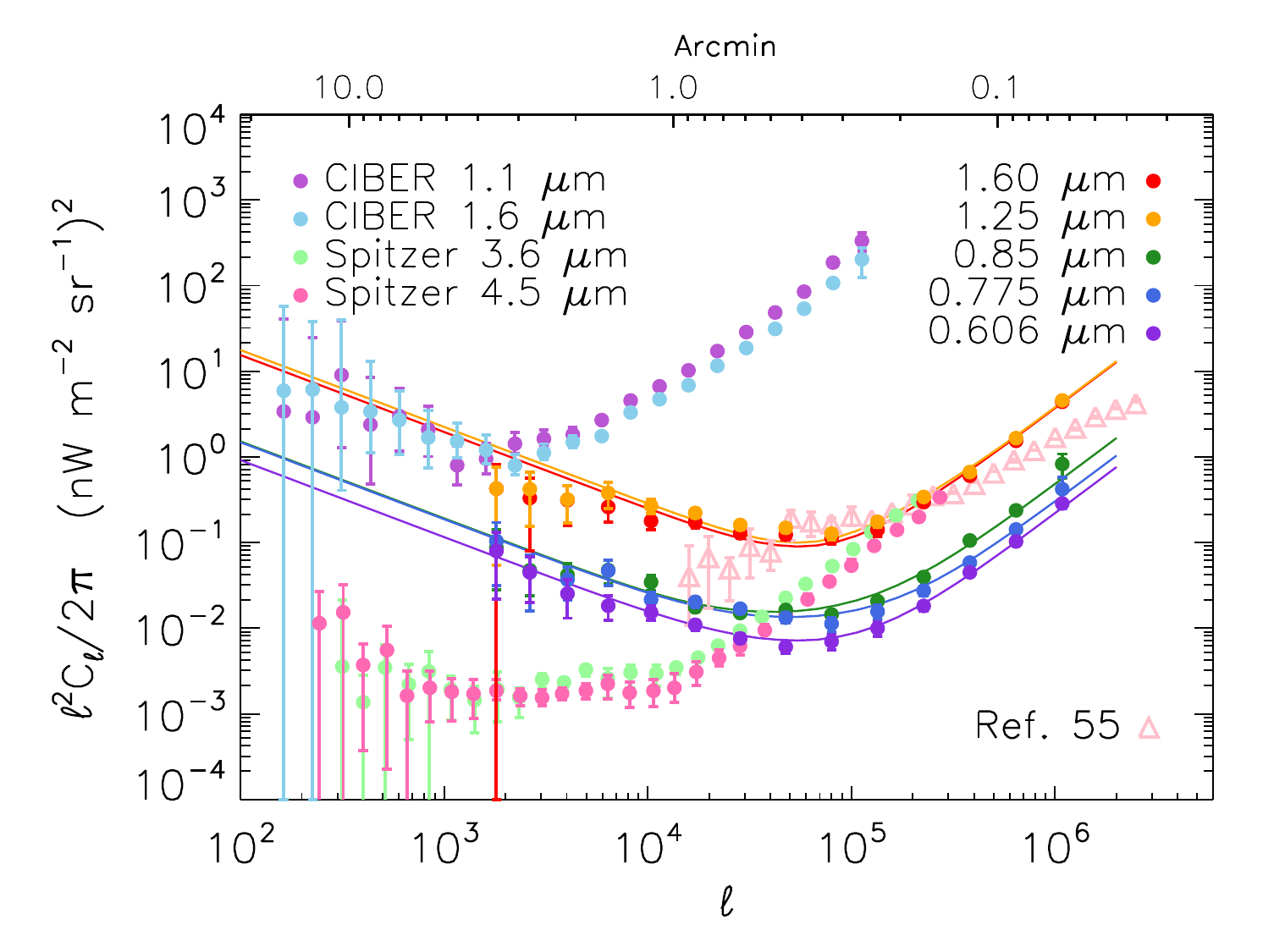}
  \includegraphics[scale=.5]{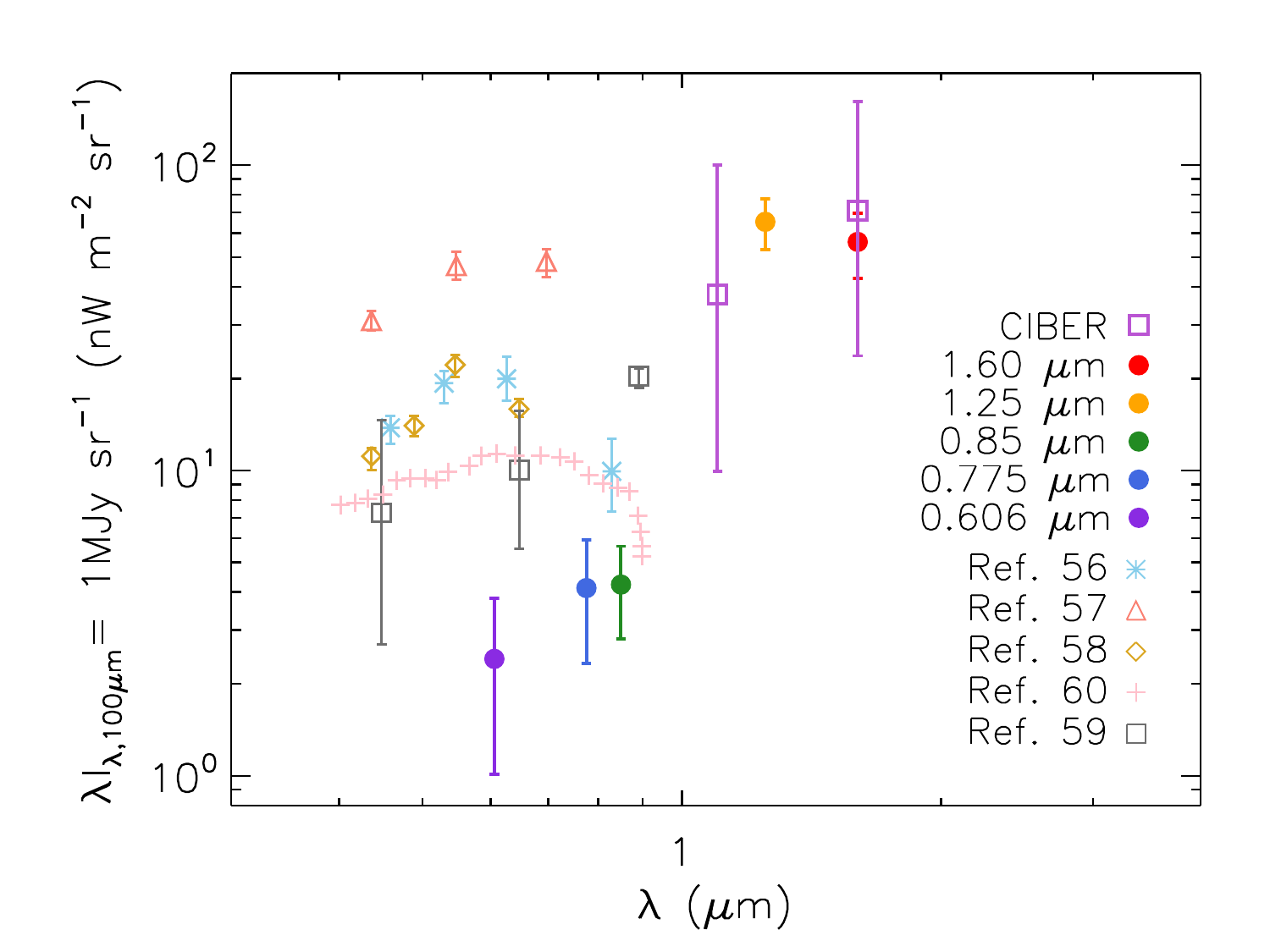}

  {\bf Figure 5 $|$ Various auto-spectra and the spectral energy distribution of diffuse galactic light.} {\bf Left,} 
  CANDELS corrected power spectra plotted against the CIBER\cite{Zemcov2014}, Spitzer\cite{Cooray2012b} and 
  NICMOS\cite{Thompson2007} measurements (see also  Ref.~\cite{Donnerstein15} for more recent NICMOS measurements). 
  The power spectrum resulting from the NICMOS analysis was measured from a MultiDrizzle map and has not been corrected for the 
  transfer function and mode-coupling matrix resulting from source masking as 
  discussed in our Methods section. Therefore we show it as a comparison but do not use it in our modeling.  
  The error bars are 1~$\sigma$ uncertainties that account for all sources of noise and error,
  including map-making, calibration, detector noise, and cosmic variance associated with finite size of the survey.
  {\bf Right:} Optical and infrared diffuse galactic light (DGL) spectrum. The CANDELS points are taken from the DGL model components at 
  $10^4\, \le \ell \le \,3\times 10^4$, and the CIBER points are taken directly from Fig~2. of Ref.\cite{Zemcov2014} where 
  they subtract off the shot noise component from their data. The galactic latitude for the optical points are 
  $|b|\,\simeq 39^{\circ}, 32^{\circ}, 41^{\circ}, 40^{\circ}$ for the points labeled Witt\cite{Witt2008}, Paley\cite{Paley1991}, 
  Ienaka\cite{Ienaka2013} and Guhathakurta\cite{Guh1989}. The Brandt\cite{Brandt2012} points are modeled over the full sky. 
  GOODS-S is at a galactic latitude of $|b| = 54^{\circ}$.
\end{figure}

\clearpage

\begin{figure}
  \begin{center}
    \includegraphics[scale=.9]{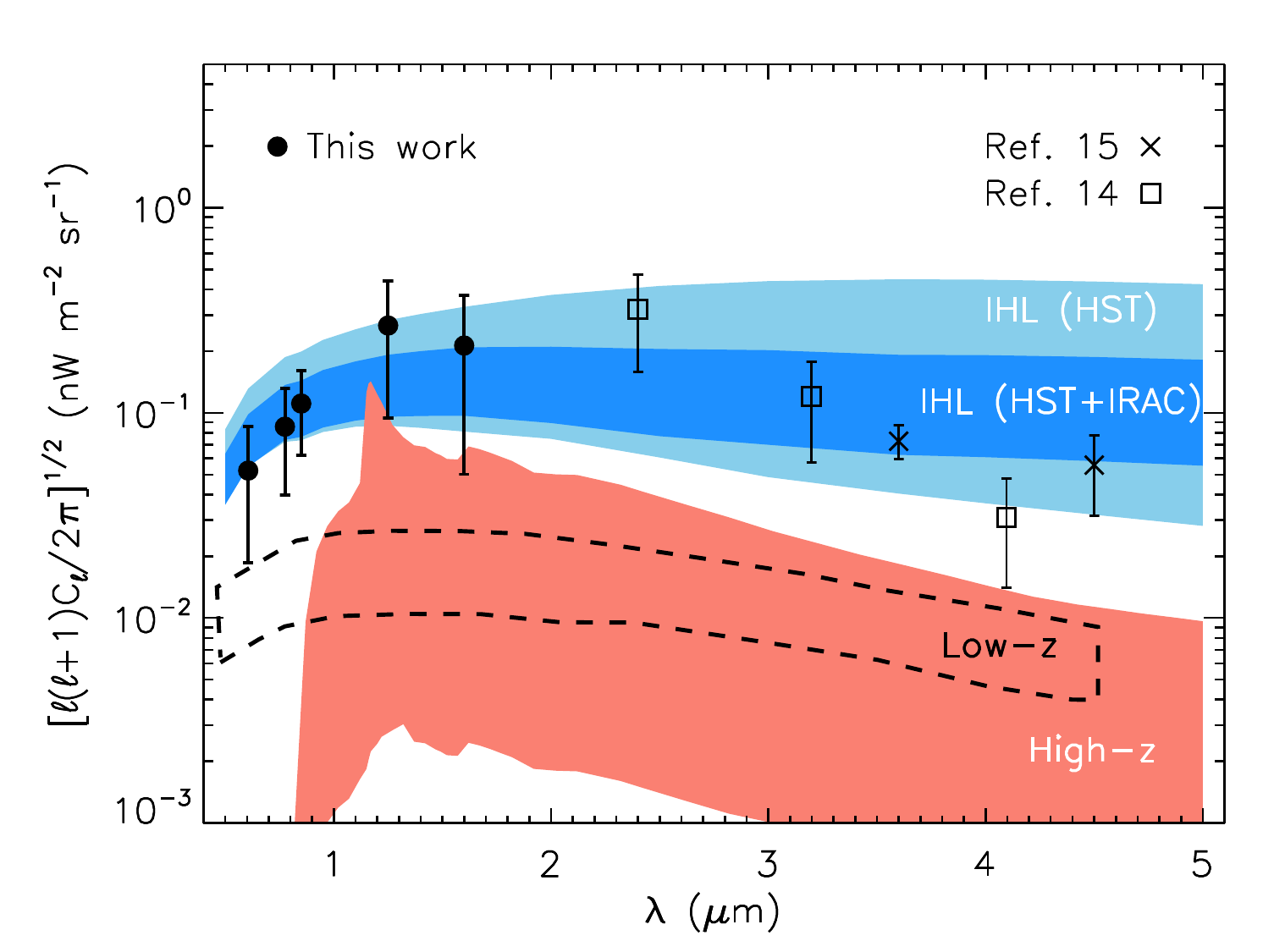}
  \end{center}
  {\bf Figure 6 $|$ Spectral energy distribution of optical and infrared fluctuations at arcminute angular scale.} The 
  Hubble/CANDELS points are averaged over $10^4\, \le \ell \le \,3\times 10^4$, with the best-fit shot noise and 
  DGL components subtracted. Our model fits for the high-redshift and IHL components, with their 
  $1~\sigma$ bounds, are shown as the filled regions. The errors here are propagated from the errors on the auto spectrum
  at the same $\ell$ range. The light blue region shows the 1~$\sigma$ confidence bound
  for the IHL component when we use only the HST data in our model fitting; the dark blue region shows the 1~$\sigma$
  confidence bound for the IHL component when use both the HST and Spitzer IRAC data in our model fitting. The light red
  colored region signifies the 1~$\sigma$ error bound for the high-redshift model component. The dashed line corresponds
  to the 1~$\sigma$ bound for the low-redshift component. The Spitzer\cite{Cooray2012b} and 
  AKARI\cite{Matsumoto2011} data are taken from previous measurements at $\ell = 3000$. Note the 
  spectral dependence difference between the high-redshift signal and IHL. Below 0.8 $\mu$m we do not 
  expect any signal from $z > 8$ galaxies. The presence of fluctuations at optical wavelengths requires a low-redshift
  signal in addition to high-redshift sources to explain combined optical and IR background intensity fluctuations.
  \label{fig2}
\end{figure}

\clearpage

\begin{figure}
  \begin{center}
    \includegraphics[scale=1]{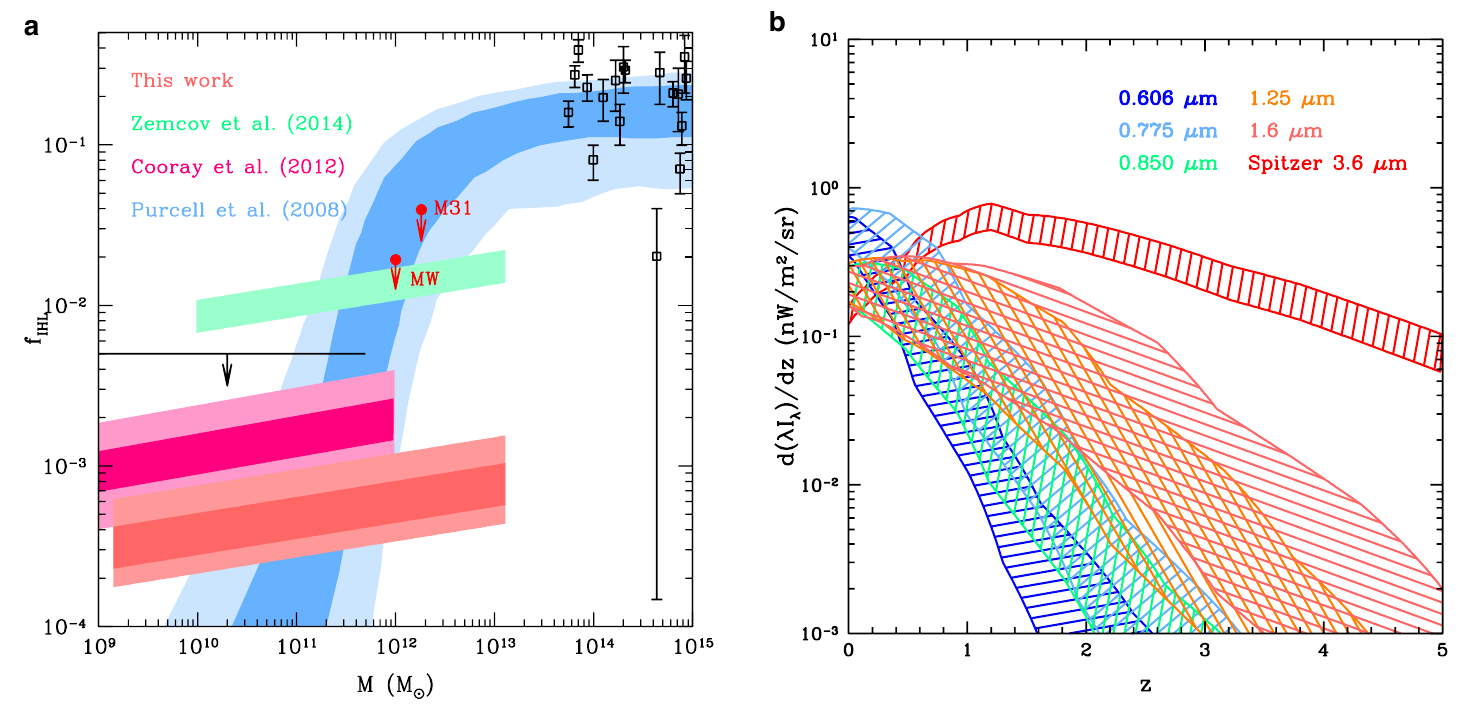}
  \end{center}

      {\bf Figure 7 $|$ Intrahalo light fraction and model intensities}
      ({\bf a}) shows $f_{\mathrm{IHL}}$, 
      the intrahalo light fraction, as a function of halo mass. The dark and light shaded regions show the 95\% and 
      68\% ranges of $f_{\mathrm{IHL}}$ from anisotropy measurements, and from an analytical prediction\cite{Purcell2007} (blue). 
      Intracluster measurements are shown as boxes\cite{Gonzalez2005}, with 1~$\sigma$ errors. The red downward arrows denote 
      the 95\% confidence upper limit on $f_{\rm IHL}$ estimated for Andromeda (M31) and our Milky Way (MW), following Figure 2
      of Ref.~\cite{Cooray2012b}.
      ({\bf b}),  $d(\lambda\,I_{\lambda})/dz$ from the model, as a function of redshift. We show the 68\% confidence uncertainties 
      derived from MCMC fitting of the data at 0.606, 0.775, 0.850, 1.25 and 1.6 $\mu$m. The total IHL intensity is 
      $0.13^{+0.08}_{-0.05}, 0.24^{+0.17}_{-0.11}, 0.28^{+0.21}_{-0.13},$ $0.45^{+0.43}_{-0.24},$ and $0.54^{+0.58}_{-0.31}$ \nw~ for 
      0.606, 0.775, 0.850, 1.25 and 1.6~$\mu$m, respectively.
\end{figure}

\clearpage

\begin{table}[!t]
\begin{center}
\begin{tabular}{l | c | c | c }
\hline\hline
 Parameter & Best fit & Best fit (no high-$z$) & Prior min, max \\
\hline
${\rm log_{10}(A_{8\le z \le 13})}$ & $1.19^{+0.27}_{-2.62} $          & -                      
             &  $-5$, 7\\
${\rm log_{10}(A_{IHL})}$         & $-3.23^{+0.14}_{-0.12}$           &  $-3.32^{+0.25}_{-0.09} 
$             &   $-$6, 10\\
$\alpha$                         & $1.00^{+0.61}_{-0.99} $           &  $1.35^{+0.39}_{-0.73} $ 
            &   $-5$, 5\\
$f_{\rm low-z}$                    & $0.47\pm 0.03 $                 &  $0.47\pm 0.03 $                  &   0.1, 10\\
$\rm A_{DGL}^{1.6}$                &$(3.74^{+0.30}_{-0.45})\times 10^{4}$ & $(3.72^{+0.35}_{-0.38})\times 10^{4}$  &  $10^3$, $10^5$\\
$\rm A_{DGL}^{1.1 \& 1.25}$         & $(4.35^{+0.54}_{-0.79})\times 10^{4}$ & $(4.72^{+0.42}_{-0.48})\times 10^{4} $ & $10^3$,$10^5$\\
$\rm A_{DGL}^{0.850}$              & $(2.83^{+0.40}_{-0.42})\times 10^{3}$ & $(2.77^{+0.32}_{-0.34})\times 10^{3}$ &  $10^2,10^4$\\
$\rm A_{DGL}^{0.775}$              & $(2.74^{-0.36}_{-0.38})\times 10^{3}$ & $(2.65^{+0.38}_{-0.48})\times 10^{3}$  & $10^2,10^4$\\
$\rm A_{DGL}^{0.606}$              & $(1.61^{+0.20}_{-0.40})\times 10^{3}$ & $(1.43^{+0.23}_{-0.22})\times 10^{3}$   &  $10^2$, $10^4$\\
$\rm C_{\ell,shot}^{1.6}$           & $(7.54 \pm 0.13)\times 10^{-11}$  & $(7.54\pm 0.13)\times 10^{-11}$    & $10^{-11}$,$ 10^{-10}$\\
$\rm C_{\ell,shot}^{1.25}$          & $(7.77^{+0.21}_{-0.28})\times 10^{-11}$  & $(7.77\pm 0.14)\times10^{-11}$ & $10^{-11}$, $ 10^{-10}$\\
$\rm C_{\ell,shot}^{0.850}$         & $(7.73^{+0.75}_{-0.45})\times 10^{-12}$ & $(8.10\pm 0.45)\times 10^{-12}$ & $10^{-12}$,$ 10^{-11}$\\
$\rm C_{\ell,shot}^{0.775}$         & $(4.60^{+0.50}_{-0.30})\times 10^{-12}$ & $(4.65\pm 0.30)\times 10^{-12}$ & $10^{-12}$, $ 10^{-11}$\\
$\rm C_{\ell,shot}^{0.606}$         & $(3.27^{+0.24}_{-0.21})\times 10^{-12}$ & $(3.39\pm 0.15)\times 10^{-12}$ & $10^{-13}$, $10^{-11}$\\
\hline
\end{tabular}
\end{center}    
\end{table}
  {\bf Table 1 $|$ Summary of free model parameters.} The best-fit values are quoted with $1\sigma$ errors. ${\rm log_{10}(A_{8 \le z \le 13})}$ is the
  high-redshift component used to constrain the SFRD during the reionization epoch, which is fit to the 1.25 and 1.60~$\mu$m bands. 
  ${\rm log_{10}(A_{IHL})}$ and $\alpha$ are the two parameters necessary to describe the IHL component (to wit: $C_\ell^{\rm IHL}$ in Equation 9). 
  $f_{\rm low-z}$ is the low redshift scaling factor which varies the low redshift power spectrum within a 1~$\sigma$ uncertainty. A$_{\rm DGL}^i$ and
   $C_{\ell,\rm{shot}}^i$ are respectively the DGL amplitude scaling factor and shot noise at wavelength $i$. All parameter values have units of (\nw)$^2$.

\clearpage

\begin{figure}
  \includegraphics[scale=.8]{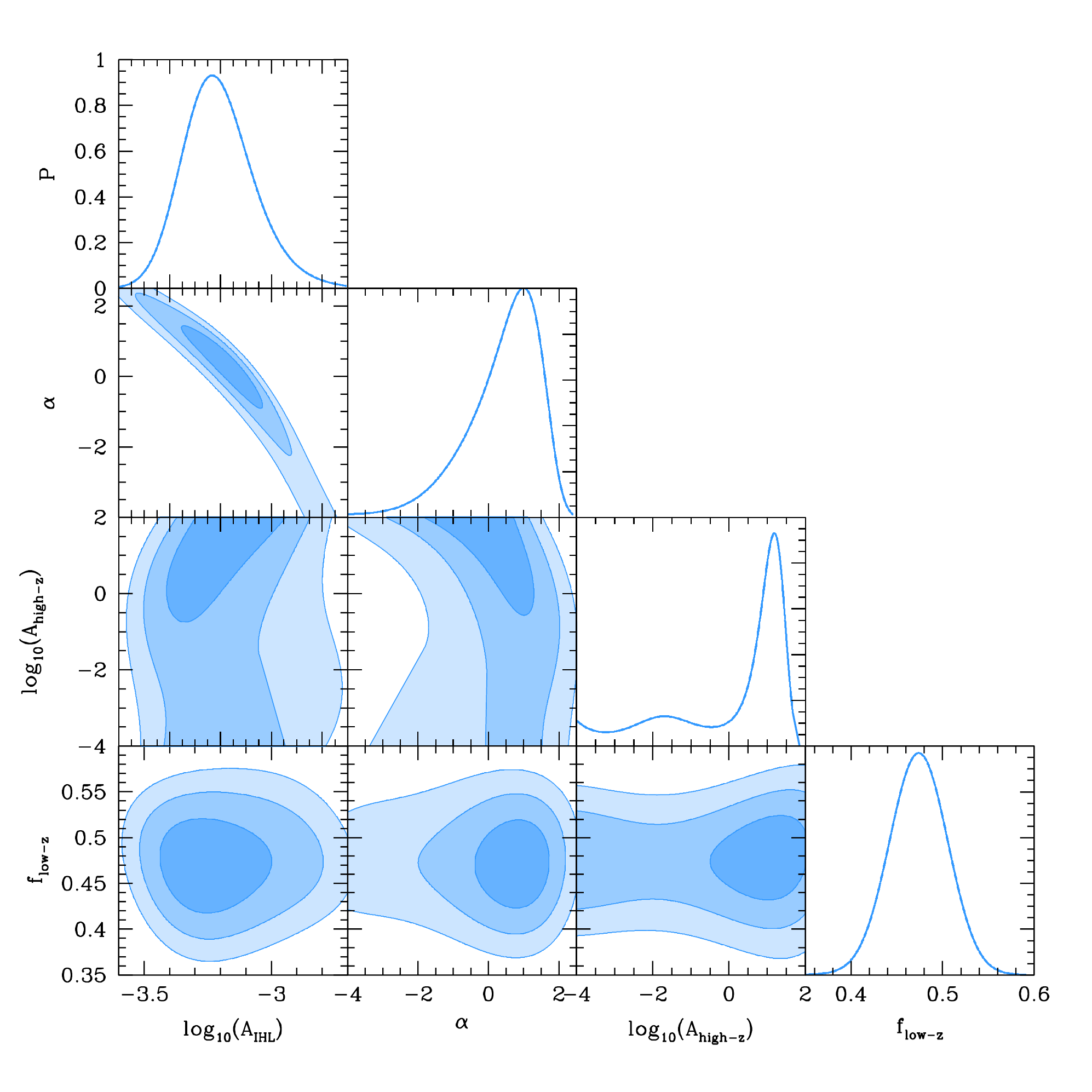}

  {\bf Figure 8 $|$ Probability distributions of fitted model parameters}. Here we show the probablity density distributions for our fitted 
  model parameters ${\rm log_{10}(A_{IHL})}$, $\alpha$, $f_{\rm low-z}$, and ${\rm log_{10}(A_{high-z})}$ corresponding to the distribution 
  from $8\le z \le 13$. The single curves on the outermost column of each row, labeled with a ``P'', show the marginalized probability 
  distribution for each parameter labeled on the bottom of the figure. Contour regions to the left of these probability distributions 
  show how the parameters scale with one another. Each of the shaded regions in the contours correspond to the 1, 2 and 3~$\sigma$ uncertainty ranges.
\end{figure}

\clearpage

\begin{figure}
  \begin{center}
    \includegraphics[scale=.7]{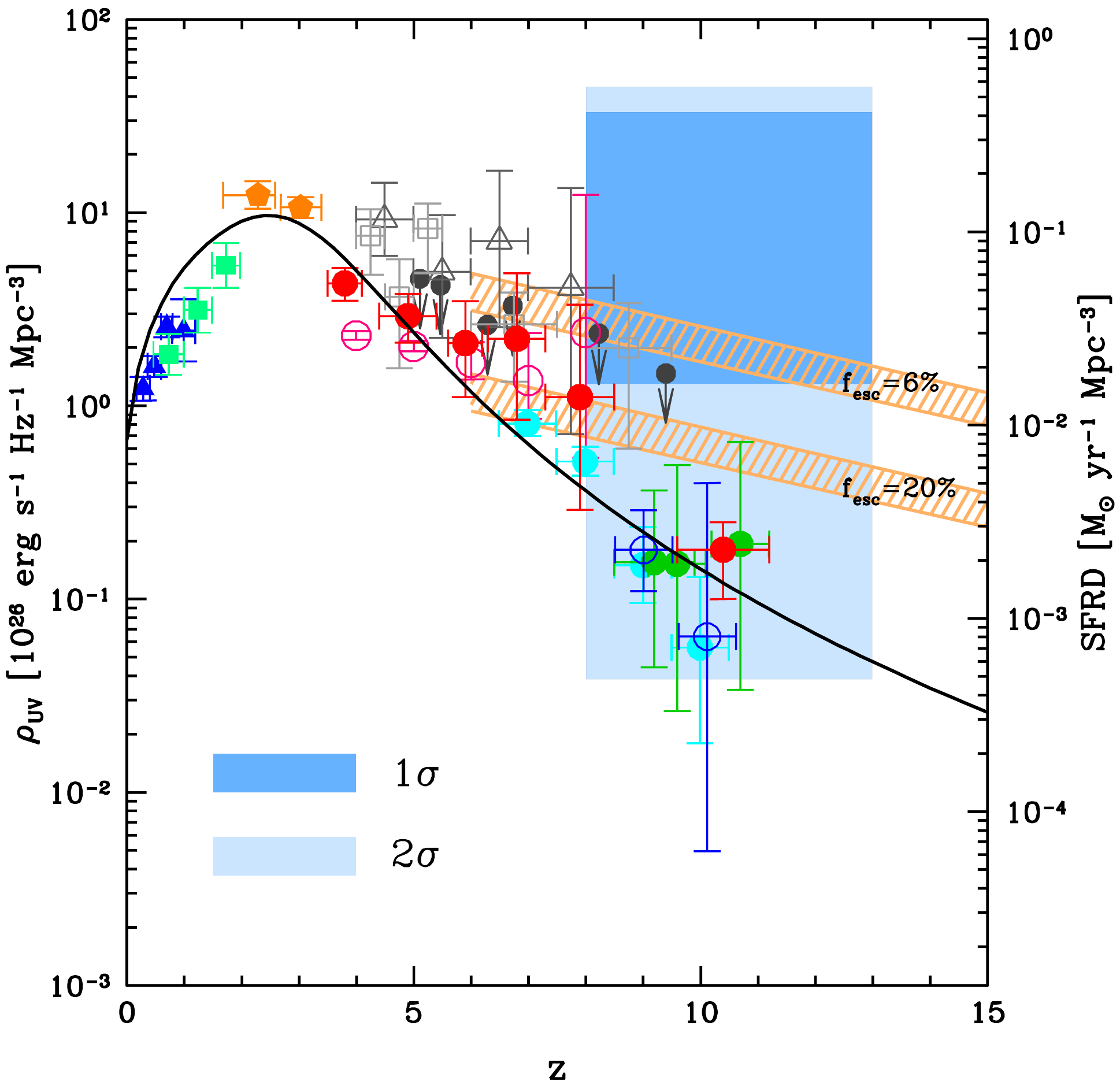}
  \end{center}

  {\bf Figure 9 $|$ The UV luminosity density and star-formation rate density as measured with intensity fluctuations}.
  Plotted here is the specific UV luminosity density (left axis), with the equivalent star formation rate density (SFRD, right axis), as a 
  function of the redshift $z$. We show the 1$~\sigma$ and 2$~\sigma$ error bounds in our redshift bin as the light and dark blue regions.
  Results from low-redshift surveys are shown as blue triangles\cite{Sch2005}, bright green squares\cite{Oesch2010}, and orange 
  pentagons\cite{Reddy2009}.  At $z \sim 4$ to 10 the star formation rate density is shown to decrease with increasing redshift 
  as measured by previous works, plotted as filled cyan circles\cite{Mclure2013}, filled red circles\cite{Bouwens2014b}, open 
  red circles\cite{Finkelstein2014} filled green circles\cite{Zheng2012,Coe2013} and open blue circles\cite{Oesch2014}. Gamma ray 
  burst (GRB) studies are shown as gray triangles\cite{Kistler2009}, squares\cite{Robertson2012} and dark gray circles\cite{Tanvir2012}. 
  Except for Ref.~\cite{Mclure2013}, other estimates are luminosity function extrapolations and integrations down to $M_{\textrm{UV}} = -13$.
  Our measured star formation rate densities are consistent with previous works between $z \sim 8$ to 10, however only extremely      
  bright galaxies are directly detected in the aforementioned works with extrapolations down to $M_{\textrm{UV}} = -13$ involves 
  the measured faint-end slope of the luminosity function. For reference we plot the theoretically expected relation\cite{Madau1999} between 
  UV luminosity density and redshift to reionize the universe and/or to maintain reionization using an optical depth to 
  reionization of $\tau=0.066 \pm 0.012$\cite{Planck2015}.  We take a gas clumping factor of $C=3$ and show two cases where the 
  escape fraction of galaxies $f_{\rm esc}$ is 6\% and 20\% (see also Ref.~\cite{Robertson2013}).
\end{figure}

\begin{figure}
\includegraphics{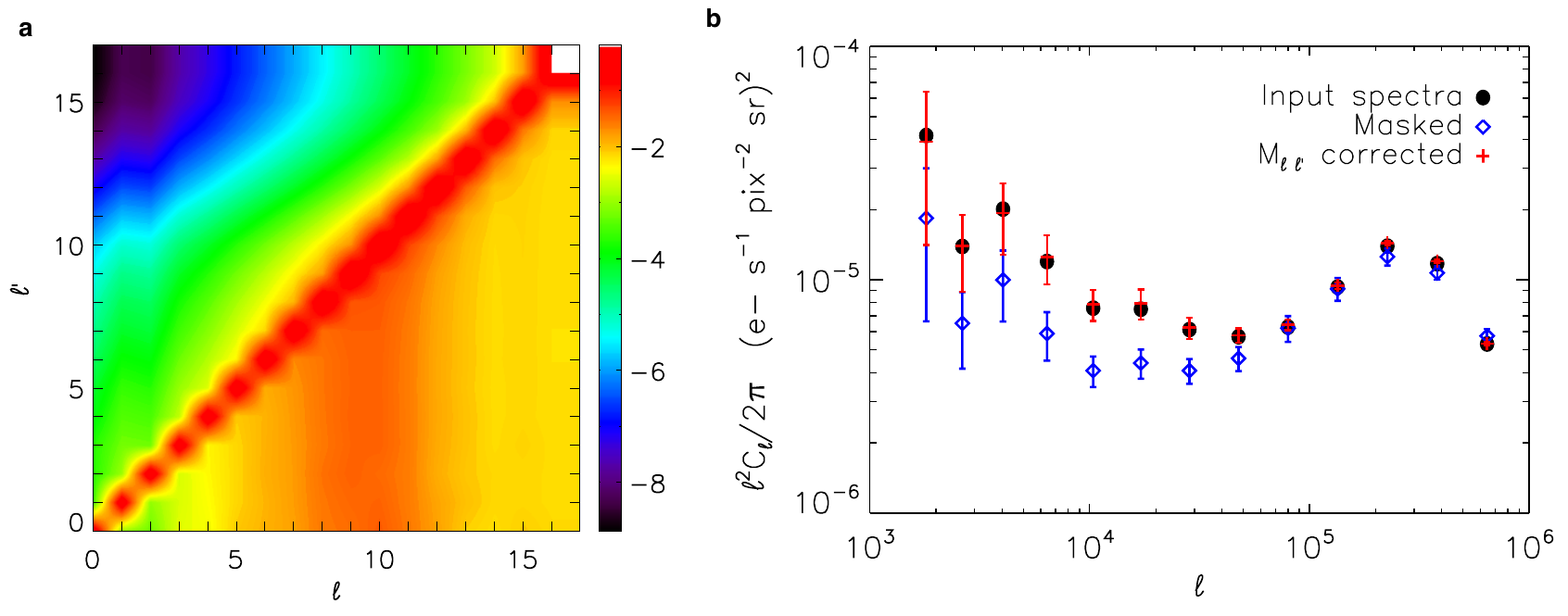}

  {\bf Supplementary Figure 1 $|$ Mode-mode coupling correction. a}, the mode-mode coupling matrix as generated 
      from our source mask. {\bf b}, \mll~validity simulation. We generate 90 Gaussian maps with a known input 
      power spectrum (black filled circles). For each realization, we apply the mask to the simulated map and compute 
      the resulting power spectrum (blue diamonds). Finally, we correct the masked power spectrum with our mode-mode 
      coupling matrix to recover the input power spectrum (red crosses), which validates our masking correction.
\end{figure}

\clearpage

\begin{figure}
\begin{center}
  \includegraphics[scale=.65]{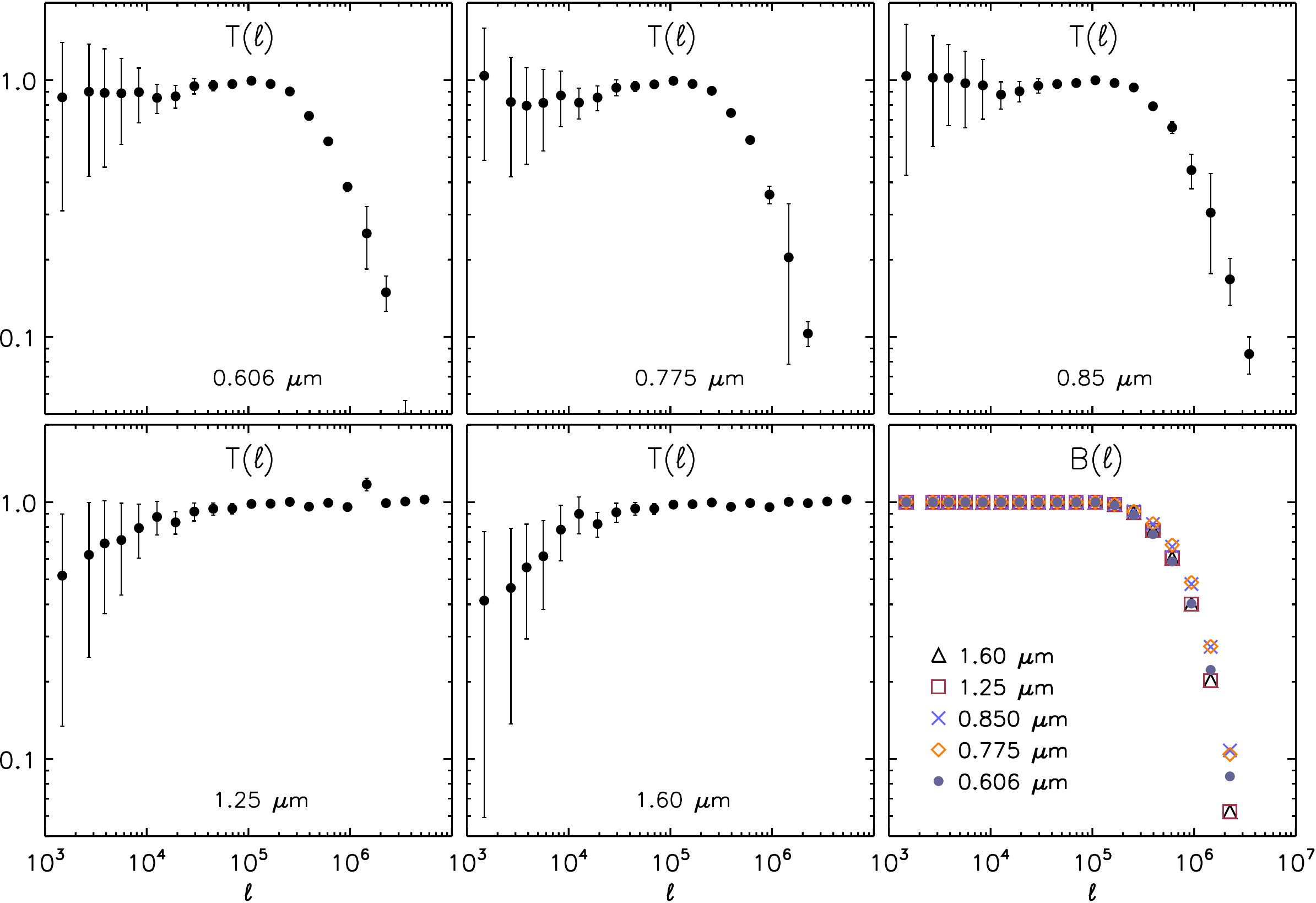}
\end{center}
    {\bf Supplementary Figure 2 $|$ Transfer and beam functions}. The T($\ell$)'s were generated from a minimum of 50 
    simulations in each band. These simulations 
    incorporate the effects of the map-making algorithm, tiling pattern, varying exposure depths, residual (temporal)
    offsets, and cropping effects, specific to each filter. The beam transfer function, $B(\ell)$, in each band is just the PSF of each 
    band in harmonic space.
\end{figure}

\clearpage


\begin{table}[!t]
\begin{center}
  \resizebox{\columnwidth}{!}{%
    \begin{tabular}{c | c | c | c | c | c}
      \hline
      $\ell$
      & \multicolumn{1}{c|}{$\ell^2~C_{\ell}^{1.6}$~/~2$\pi$}
      & \multicolumn{1}{c|}{$\ell^2~C_{\ell}^{1.25}$~/~2$\pi$}
      & \multicolumn{1}{c|}{$\ell^2~C_{\ell}^{0.850}$~/~2$\pi$}
      & \multicolumn{1}{c|}{$\ell^2~C_{\ell}^{0.775}$~/~2$\pi$}
      & \multicolumn{1}{c}{$\ell^2~C_{\ell}^{0.606}$~/~2$\pi$} \\
  \hline\hline
  $1.81\times 10^{3}$ &$ 1.10 \pm 1.17              $ & $ 1.10 \pm 0.98  $ & $ (1.86 \pm 1.33) \times 10^{-1}$ & $ (2.30 \pm 1.69) \times 10^{-1}$ & $ (1.70 \pm 1.30) \times 10^{-1}$ \\
$2.64\times 10^{3}$ &$ (8.30 \pm 6.60) \times 10^{-1}$ & $  1.08 \pm 0.72               $ & $ (9.55 \pm 5.10) \times 10^{-2}$ & $ (9.02 \pm 6.19) \times 10^{-2}$ & $ (9.01 \pm 5.38) \times 10^{-2}$ \\
$4.04\times 10^{3}$ &$ (7.78 \pm 4.09) \times 10^{-1}$ & $ (7.94 \pm 4.04) \times 10^{-1}$ & $ (8.30 \pm 3.25) \times 10^{-2}$ & $ (7.21 \pm 3.29) \times 10^{-2}$ & $ (4.74 \pm 2.47) \times 10^{-2}$ \\
$6.40\times 10^{3}$ &$ (6.33 \pm 2.30) \times 10^{-1}$ & $ (9.61 \pm 3.53) \times 10^{-1}$ & $ (9.61 \pm 3.23) \times 10^{-2}$ & $ (9.40 \pm 3.34) \times 10^{-2}$ & $ (3.29 \pm 1.17) \times 10^{-2}$ \\
$1.04\times 10^{4}$ &$ (4.18 \pm 0.94) \times 10^{-1}$ & $ (6.60 \pm 1.43) \times 10^{-1}$ & $ (6.70 \pm 1.47) \times 10^{-2}$ & $ (3.98 \pm 0.91) \times 10^{-2}$ & $ (2.71 \pm 0.57) \times 10^{-2}$ \\
$1.71\times 10^{4}$ &$ (4.01 \pm 0.63) \times 10^{-1}$ & $ (5.29 \pm 0.77) \times 10^{-1}$ & $ (3.13 \pm 0.41) \times 10^{-2}$ & $ (3.70 \pm 0.57) \times 10^{-2}$ & $ (1.86 \pm 0.27) \times 10^{-2}$ \\
$2.85\times 10^{4}$ &$ (2.89 \pm 0.39) \times 10^{-1}$ & $ (3.71 \pm 0.46) \times 10^{-1}$ & $ (2.66 \pm 0.28) \times 10^{-2}$ & $ (3.02 \pm 0.39) \times 10^{-2}$ & $ (1.24 \pm 0.17) \times 10^{-2}$ \\
$4.76\times 10^{4}$ &$ (2.76 \pm 0.37) \times 10^{-1}$ & $ (3.39 \pm 0.41) \times 10^{-1}$ & $ (2.90 \pm 0.27) \times 10^{-2}$ & $ (2.30 \pm 0.34) \times 10^{-2}$ & $ (9.62 \pm 1.74) \times 10^{-3}$ \\
$7.99\times 10^{4}$ &$ (2.50 \pm 0.43) \times 10^{-1}$ & $ (2.85 \pm 0.46) \times 10^{-1}$ & $ (2.54 \pm 0.30) \times 10^{-2}$ & $ (1.92 \pm 0.40) \times 10^{-2}$ & $ (1.13 \pm 0.24) \times 10^{-2}$ \\
$1.34\times 10^{5}$ &$ (3.19 \pm 0.56) \times 10^{-1}$ & $ (4.03 \pm 0.63) \times 10^{-1}$ & $ (3.76 \pm 0.45) \times 10^{-2}$ & $ (2.75 \pm 0.55) \times 10^{-2}$ & $ (1.67 \pm 0.34) \times 10^{-2}$ \\
$2.26\times 10^{5}$ &$ (7.29 \pm 0.80) \times 10^{-1}$ & $ (8.50 \pm 0.89) \times 10^{-1}$ & $ (7.76 \pm 0.69) \times 10^{-2}$ & $ (5.16 \pm 0.74) \times 10^{-2}$ & $ (3.27 \pm 0.47) \times 10^{-2}$ \\
$3.82\times 10^{5}$ &$  1.62 \pm 0.09               $ & $  1.80 \pm 0.11                $ & $ (2.33 \pm 0.13) \times 10^{-1}$ & $ (1.20 \pm 0.10) \times 10^{-1}$ & $ (8.96 \pm 0.75) \times 10^{-2}$ \\
$6.44\times 10^{5}$ &$  4.59 \pm 0.13  $               & $  4.96 \pm 0.16                $ & $ (5.73 \pm 0.48) \times 10^{-1}$ & $ (3.24 \pm 0.22) \times 10^{-1}$ & $ (2.26 \pm 0.13) \times 10^{-1}$ \\
$1.09\times 10^{6}$ &$ (1.49 \pm 0.02) \times 10^{1}$ & $ (1.55 \pm 0.05) \times 10^{1}$ & $  2.31 \pm 0.80                $ & $  1.09 \pm 0.47               $ & $ (7.08 \pm 0.70) \times 10^{-1}$ \\
\hline
    \end{tabular}%
    }
\end{center}
{\bf Supplementary Table 1 $|$ Final HST power spectra.} Corrected auto-spectra, $\ell^2C_{\ell}$/2$\pi$ in units of (\nw)$^2$, for 5 bands. The quoted errors are the 1~$\sigma$ uncertainties. 
\end{table}

\clearpage
\begin{figure}
  {\bf Supplementary Note 1 $|$ Data Availability:} The self-calibrated mosaics used for the fluctuation study, including jack-knives with data separated to epochs, and the detected source mask, mode-coupling matrix, beam functions, and the transfer functions are available at http://herschel.uci.edu/CANDELS
\end{figure}

\end{document}